\documentclass{article}

\usepackage[margin=1in]{geometry}
\usepackage{subfigure}
\usepackage{graphicx}
\usepackage{color}
\usepackage{url,booktabs} 
\usepackage{mdwlist}
\usepackage{pslatex}
\usepackage{amsmath}
\usepackage{amssymb}
\usepackage{multirow}

\title{Unlinking super-linkers: the topology of epidemic response (Covid-19)}

\author{
{Shishir Nagaraja}\\
University of Strathclyde\\
\tt{shishir.nagaraja@strath.ac.uk}
}

\begin{document}

\maketitle

\newcommand{\paragraphb}[1]{\vspace{0.03in}\noindent{\bf #1} }
\newcommand{\paragraphe}[1]{\vspace{0.03in}\noindent{\em #1} }
\newcommand{\paragraphbe}[1]{\vspace{0.03in}\noindent{\bf \em #1} }

\newcommand{\scomment}{\textcolor{blue}}
\newcommand{\ecomment}{\textcolor{red}}

\begin{abstract}
  A key characteristic of the spread of infectious diseases is their ability to use efficient transmission paths within contact graphs. This enables the pathogen to maximise infection rates and spread within a target population. In this work, we devise techniques to localise infections and decrease infection rates based on a principled analysis of disease transmission paths within human-contact networks (proximity graphs). Experimental results of disease spreading shows that that at low visibility rates contact tracing slows disease spreading. However to {\em stop} disease spreading, contact tracing   requires both significant visibility (at least 60\%) into the proximity graph and the ability to place half of the population under isolation. We find that pro-actively isolating super-links --- key proximity encounters -- has significant benefits --- targeted isolation of a fourth of the population based on 35\% visibility into the proximity graph prevents an epidemic outbreak. It turns out that isolating super-spreaders is more effective than contact tracing and testing but less effective than targeting super-links. We highlight the important role of topology in epidemic outbreaks. We argue that proactive innoculation of a population by disabling super-links and super-spreaders may have an important complimentary role alongside contact tracing and testing as part of a sophisticated public-health response to epidemic outbreaks.
\end{abstract}

\section{Introduction}

There has been rapid progress in understanding how infectious diseases
spread organically, how the disease growth affects topology, and how
the topology in turn affects disease interventions and the number of
infected individuals. There is now substantial literature on the
topic: see literature surveys~\cite{eichner2003Oxford, ma2004Bejing}
and book length introduction to disease
propagation~\cite{anderson1992Oxford, daley1999Cambridge,
  diekmann2000Wiley}.

Early work modeled outbreaks of infectious diseases as a randomised
stochastic process where every possible contact between an infected
and a susceptible individual was equally likely~\cite{guo2016PLoS,
  guo2019BMC, shubin2016PLoS}.  These models assume a disease
transmission network that is modeled by an Erd\"os-Renyi
graph~\cite{erdos1960MathInstitute}, which is mathematically
interesting but does not model real-world contact patterns
accurately. In real networks, the infection chain may consist of
individuals who are likely to be pairwise acquainted. In the case of a
respiratory disease like Covid-19, for successful transmission, two
individuals must be in contact that is sufficiently close enough to
enable the transmission of the respiratory
disease~\cite{ferretti2020Science}. Thus it is likely that many
successful transmissions involve pairwise acquaintances rather than
contact between random individuals. In other words, while the random
contact model is a good first approximation, in real networks the
path-lengths are likely to be a lot smaller leading to faster disease
transmission than modeled previously. This is known as the
'small-world' concept and it was first popularised in the context of
social networks by the sociologist Stanley Milgram in
1967~\cite{M67}. Indeed existing epidemiology models use the {\em
  reproduction number} $R_0$ parameter as central component of their
framework. $R_0$ represents the average number of infections caused by
an infected person, and is assumed to be distributed according to a
Normal distribution (assuming a purely random contact contact
pattern).

Connectivity plays a significant role in the spread of infectious
diseases. Epidemiologists and medical practitioners expend much effort
in tracking down people who were in the proximity of an infected
person, followed by isolating them, and testing them. Techniques such
as proactive screening of the population and testing random samples
for pathogens all require significant effort. Additionally, these
measures are said to work best when dealing with local clusters of
infection. In particular, they don't scale to protect large
populations especially against a virus like SARS-CoV-2 that has a
significant incubation period of 7 days~\cite{backer2020ECDPC}. When
diseases start spreading over networks of inter-personal contacts,
responders can rapidly lose control of the situation. In the absence
of appropriate interventions the numbers of infections and deaths
rapidly increase and responders may be left with no other option but
to isolate the entire population (lockdown). Aside from the severe
economic and social repercussions, a lockdown does not stop the
disease as transmission resumes soon after the suspension is
lifted. As such social lockdowns are a one-off emergency measure that
cannot be repeatedly applied due to the volatility they induce within
human societies.

%

Targeted interventions can be much more effective and efficient. The
intuition is that a small number of key individuals are believed to
play a major role in disease transmission. Indeed many countries have
successfully disrupted the spread of HIV/AIDS by targeting sex
workers~\cite{Thom03}. In contrast the efforts to contain Covid-19
have been largely restricted to the deployment of contact tracing
combined with testing and
isolation~\cite{culnane2020uk,bay2020bluetrace,gov2020privacy,wiertz2020predicted,cdc2020principles,yoneki2012flu,dp3t}. As
an disease response strategy this is fairly ``leaky'' -- even if a
fraction of contacts remain untraced or unknown, then the disease
continues to spread picking up pace soon after. As such ``leakages'' are
hard to detect and expensive to contain.  To automate contact tracing,
apps based on pseudonymous Bluetooth beacons are at a stage of early
deployment in many countries. However they require at least 60\% of
the population to adopt it in order to gain control of the
disease. Current reports~\cite{economist2020economist} indicate the
adoption rates to be in the range of between 12\% to 40\%, crucially
limiting the efficacy of a targeted strategy based on
contact-tracing. There are a number of example contact tracing
applications, for example bluetrace \cite{bay2020bluetrace} and
\cite{culnane2020uk}.

In this study, we systematically analyse various elements of human
communication structures to develop sophisticated {\em proactive
  isolation strategies} that compliment contact tracing
approaches. Is it possible, for example, to slow down or entirely stop
the spread of an epidemic by pro-actively isolating (or quarantining)
key individuals who play a vital role in disease spreading due to
their positionality in the contact network. Is it possible that a low
vertex-order node who has contacts within different groups of people
who may otherwise have no contact with each other should be encouraged
to self-isolate? There is loose precedent in epidemiology literature
which discusses the concept of {\em super-spreaders} defined as high
vertex-order nodes play a disproportionate role in disease
spreading. However such nodes may simply be individuals with essential
roles such as a bus driver or a nurse, who happen to interact with a
large number of people within a short span of time. We answer key
research questions: Is it possible to identify key individuals that
must pro-actively self-isolate using the contact graph alone? Is it
possible to identify such individuals even at low rates of app
adoption? and, how to achieve this in a privacy-preserving manner,
moving away from extremes such as undertaking pan-societal lockdown
measures.

%

\section{Evaluation framework for measuring disease propagation}

 The focus of our work is on the probability that a susceptible
 individual will become an infected individual. Given an initial
 infection source within a population, we wish to be able to determine
 transmission chains -- who infected whom -- leading to the
 requirement to identify both infectious and susceptible
 individuals. The determination of infected or infectious individuals
 is driven by the probability that a specific individual is the
 originator of a given (infection) transmission chain. Similarly,
 identifying the set of susceptible individuals who are potentially at
 risk from an infection chain originating at a certain individual is
 also an important requirement to combat an infectious disease.

The objective of our analysis is to determine how the topology of
disease propagation affects the efforts of public health responders in
identifying the endpoints of an transmission chain using contact tracing
approaches. The effectiveness of such interventions depends on the
topology of the propagation network. A disease intervention
mechanism should help responders. If responders are unable to narrow
down on susceptible individuals at risk of joining an
infection chain beyond their initial knowledge prior to the
intervention, then the propagation network can be described as being
resistant to an intervention strategy.

We are interested in the effects of topology alone and ignore the
effects of vaccination, acquired immunity, and changing rates of
infectiousness on disease propagation.


\subsection{Measuring disease }
There are several ways to measure disease propagation. Conventional
approaches typically use the counts of susceptible, infected,
infectious, and recovered individuals over time. These counts help
understand the current status of disease propagation but provide no
understanding of the network's (future) {\em capacity for disease
  propagation}. If we think of the network as a communication channel
-- where the efficiency of disease spreading (communication) is a
function of channel (network topology) efficiency -- then we can see
that tools from Shannon theory of communications, specifically the
capacity measures can be used to better understand disease
transmission.

We therefore introduce a new metric, called the disease transmission
potential of a network defined as the quantum of disease transmission
that can occur due to a single infection occurring anywhere within the
network. This is mathematically represented as the entropy of the
probability distribution of contracting an infection over all
individuals within a population. This can also be equivalently
understood as the amount of information the public-health responder is
missing in order to identify the susceptible individuals that are most
at risk of being infected. Low  potential means there
isn't much work remaining to be done in terms of identifying
infectious individuals and isolating them from the susceptible
population. On the other hand, high  potential implies the
responder has little knowledge and therefore high uncertainty about
what network elements to disable in order to stop transmission.

\begin{equation}
  \mathcal{I}=\mathcal{E}[\beta_{i}] = \frac {-
  \sum_{i}Pr[\beta_{i}]log_{2}Pr[\beta_{i}]}{log_{2} N}
  \label{eqn-entropy1}
\end{equation}

Transmission potential of a network is highest when all individuals
within a population are at their maximal likelihood of contracting the
disease. The maximum potential of a network of $N$ individuals is
$\frac{log_2 N}{log_2 N}$ and the minimum is zero. Thus we can define
a health response to have maximal efficiency when the potential of the
propagation network is zero. This the upper bound of efficiency for a
response strategy.

\subsection{Measuring disease propagation}
\label{metrics}

\paragraphb{Secondary transmission potential}
In analyzing disease propagation on a particular network topology
we need to examine the probability that a specific individual is part
of an infection chain originating at a particular node at a certain
time. In order to link the infection of a susceptible node to a
certain originator, the responder must trace those in the proximity of
the infectious originator and work their way through multiple
infection chains until there are no further infected individuals
linked to the origin. Let the disease propagation network be a
directed graph $G(V, E)$. If node $i$ gets infected in an disease
transmission event $s_{ij}$, and is part of an infection chain that
ends at $j$, then for a event $s_{k}^t$ where node $k$ gets infected
at time $t$, the responder must link infection event $s_{k}^{t}$ to
$s_{ij}$.

Applying the transmission potential metric, we have:
\[ \mathcal{I}=\mathcal{E}(p_{ij})\] where \(p_{ij}=Pr[s_{k}^{t}\ is\ s_{ij}] \)
 is the probability distribution over all the susceptible nodes in $V$.

Suppose the disease is seeded by an infection event through a randomly
chosen susceptible node within the network. Then after an infinite
number of steps, the probability that a randomly chosen individual in
the network is infected is given by stationary distribution of the
Markov chain $\pi$.  Let $q^{(0)}$ be the initial probability
distribution over individuals where the pathogen is introduced into
the network, this is equivalent to the distribution of initial viral
load across susceptible individuals.  $q^{(t)}$ then, is the
probability distribution of infected nodes at which the pathogen is
present after $t$ steps. (this is also known as the state probability
vector of the Markov chain at time $t \geq 0$). With increasing $t$
one would like to see that $q^{(t)}$ merges with $\pi$. The rate at
which this takes place is known as the {\em convergence rate} of the
Markov chain, and the difference itself is called the {\em relative
  point-wise distance} defined as:
\begin{equation}
\Delta(t) = max_{i} \frac{|q_{i}^{t}-\pi_{i}|}{\pi_{i}}
\label{mc-delta-qpi}
\end{equation}

The smaller the relative point-wise distance, faster the convergence,
and faster the disease transmission within a network. It is now easy
to see that the maximum transmission potential \(Pr [x=infected |
  y=infectious\ originator] \) the network can provide is the entropy
of the stationary distribution of the chain.

\begin{equation}
\mathcal{I}_{network}=\mathcal{E}(\pi)
\label{eqn-max-anon}
\end{equation}.

%

When $P$ is the transition matrix of the chain it is well known that
$P$ has $n$ real eigen-vectors $\pi_{i}$ and $n$ eigenvalues
$\lambda_{i}$ \cite{W01}. By using the relation
\(q^{(t)}=q^{(0)}P^{(t)}\), we calculate the probability distribution
of a node being infected after having faced a viral load from an
infection chain of length $t$.

\paragraph{Primary or origin transmission potential:}
Next, we consider the probability distribution of potential
originators of an infection outbreak. This may also be modeled by a
Markovian random walk. For a destination node $y$, consider all random
walks terminating at $y$. In order to achieve maximal potential, all
these infection chains must be long enough for the respective state
probability vector to converge with the stationary distribution. Since
this applies equally to all originator nodes in the network, the
primary transmission potential is given by:
\[ Pr[X=x | y] = \frac{1}{N=|V|}\].

\subsection{Modeling disease spreading as a random walk}
To simulate epidemic disease spreading on a network, we use Markov
chains. This is a stochastic process that closely matches the way the
disease spreads from an infectious individual to a susceptible
individual.

The process of disease spreading wherein a single individual is the
source of infection within a population, is equivalent to first
selecting a random individual and then one or more random
(susceptible) neighbours of the first (infectious) individual,
repeating this process until there are no more susceptible
individuals. Hence we may model disease spreading as a random walk on
the disease propagation graph, with individuals being the states of
the Markov chain process.

Aside from graph topology there are other factors that affect the
dynamics of an outbreak. The transmission potential of the disease
from an infected to a susceptible member depends on the infectivity
of the disease. Further, the infectivity of the individual (i.e the
progression of the infection within the carrier) varies over
time~\cite{ratsitorahina2000Lancet} and is often assumed to follow a
Weibull distribution~\cite{ferreti2020AAAS, lessler2009NEJM}. We will
focus on the impact of the topology of the disease transmission
network in this study and ignore the effects of incubation period,
generation period, symptom duration which have been studied
elsewhere~\cite{kaihao2020elsevier,ferreti2020AAAS,guo2019BMC}.

\section{Public Health Interventions}

Having discussed the role of network topology in disease propagation,
we now discuss interventions. The responder's goal is to
deploy passive methods such as disease surveillance and active methods
such as quarantining or recommending self-isolation of individuals
within the proximity network. It is natural to ask how these measures
can be applied within a public-health response to achieve the best outcomes.

Intuitively, we can answer this question by stating that the responder
should issue isolation/quarantine instructions according to a strategy
that maximises the probability that the outbreak is ``intercepted''
regardless of the specific path followed by the chain of disease
transmission. We start with the observation that there are principally
two approaches towards achieving an optimal result: vertex-based {\em
  super-spreader} isolation or edge-based {\em super-link} isolation
approaches.

\subsection{Super-spreader isolation}
In vertex-based approaches, the goal is to locate and isolate
individuals that play a disproportionate role in disease spreading i.e
pursue {\em super-spreader} detection. As one possible definition,
super-spreaders are high vertex-order (high degree) nodes that
influence disease propagation due to the numbers of susceptible
individuals they interact with. Vertex-order nodes~\cite{N03a} are
special nodes that owing to their position in the network topology
broker large amounts of proximity encounters. The risk of fully
visible transmission chains is significantly higher in a topology
where hubs only connect to other hubs, and are responsible for a
majority of the network's proximity encounters. If disease responders
can locate and strategically target individuals that play the role of
a hub (super-spreaders) or host a weak-tie, then the percentage of
transmission chains that start and end within the subset of
contact-tracing app users can be significant. This property is known
as assortativity~\cite{N03b}, defined as the affinity of a node to
link to others that are similar or different in some way.

\paragraphb{Vertex-order super-spreaders:} In a centralised approach,
vertex order is computed by adding up the number of neighbours for
each node. To estimate vertex-order in the decentralised case, we use
long random walks. A long random walk of length $O (Log_{d} N)$ steps
where $d$ is the average degree and $N=|V|$ is the population size is
used to compute the degree distribution on the proximity graph. The
stationary distribution of a proximity graph is its degree
distribution. A random walk that is long enough ($O (Log N)$) to
achieve stationarity, results in a probability distribution over the
nodes of the graph which is also its degree distribution. Thus detecting vertex-order super-spreaders in a decentralised manner is the same as computing the result of a long random walk over a distributed graph. We provide a concrete privacy-preserving mechanism for this task in Section~\ref{sec:ppgraph}.

%
%
%

\subsection{Super-link isolation}
A complimentary approach is to isolate proximity links (graph edge)
between certain individuals that play an important role in
transmission. {\em Super-links}~\cite{G73} are edges or proximity
links that are responsible for significantly reducing average
path-lengths in networks of tightly knit communities such as social
networks. This is important in the context of community transmission
of an infectious disease. Super-links inter-connect dense clusters of
nodes. Consequently disabling a super-link seems a natural defense against
disease spreading, yet this has not been investigated before. Better still, disabling a set of $k$ links (a $k$
cut) by preventing those people from getting in close proximity
partitions the graph and stops the spread of the disease with high
probability. Here's why. Consider a $k$-cut that partitions a graph
into several components of similar size. The likelihood of a node on
the tour belonging to the same partition as the predecessor node is
denoted by $p$, so with $1-p$ the next node on the tour could belong
to a different partition. The likelihood of a tour comprising all $m$
nodes from the same partition is then given by $f_{p}(j,m)$ where $j$
is the number of consecutive infected nodes already in the same
partition is defined as: \( f_{p}(j, m) = pf_p(j+1,m) \). As $m$
increases, \( \lim_{m} f_p(0,m) \rightarrow 0 \) for $0 < p < 1$. As a
concrete example, for $p=0.5$, the likelihood of a $15$-step
transmission chain not traversing a mincut, is less than $1$ per $65$
per million transmission chains (regardless of graph size), decreasing
to almost zero per million transmission chains of $20$ steps. Having
established that any path taken by the pathogen will include at least
one edge of the mincut, with probability close to $1$, what does that
mean for spreading efficiency? The public-health responder can
leverage this understanding to construct a highly effective
intervention strategy that maximises her utility regardless of the
spreading dynamics --- to block all transmission paths by placing
monitoring or quarantine orders to disable transmission along every
edge of the $k$-mincut. Further, the $k$-mincut presents a theoretical
limit on the number of disjoint transmission chains available to the
pathogen, as any chain must involve at least one edge from the
mincut. Note that this includes paths that are not shortest paths
between a pair of consecutive nodes on the proximity graph.

%

%

The basic idea is to partition the affected population into multiple
disjoint subsets of individuals that are related by a small (minimal)
number of proximity edges, by controlling the edges we hope to control
the outbreak. The approach taken by most of community detection literature~\cite{cdc2020principles}.
%
%
%
%
%

The approach we use in this paper is a combination of short and long
random walks to find efficient partitioning cuts. We do this by
applying the Botgrep algorithm~\cite{nagaraja2010botgrep} which uses
short random walks to instrument mixing time within a partition and to
minimise the leakage of walks starting from a partition. Subsequently,
Botgrep applies the probabilistic model from SybilInfer to isolate
edges which delineate the subset of the graph where mixing speed
changes. Thus the output of Botgrep is various graph subsets with
different mixing characteristics. While conventional random walks use
only the source to determine the probability of edge transition,
Botgrep uses a special probability transition matrix to implement the
random walks, where the transition probability between adjacent nodes
${i,j} \in V$ is $min (\frac{1}{d_i},\frac{1}{d_j})$, as opposed to
$\frac{1}{d_i}$ from $i$ to $j$ in Markovian random walks, where $d_i$
is the degree of node $i$. The intuition here is that proximity
networks are partitioned by inter-connected islands of proximity
encounters, where each island has a contact pattern that is different
to others. 

 Indeed, the motivation for Botgrep based random-walk partitioning is
 that the propagation network may not necessarily consist of dense
 clusters of infections separated by small cuts. While a
 {\em small-cut} is a useful theoretical starting point, propagation
 networks may not necessarily contain small-cuts that partition the
 graph into two or more components that are non-trivial in size. Thus
 a complimentary approach to small-cut detection is offered by the
 {\em Botgrep} partitioning which combines machine learning with
 random-walks to identify sub-graphs with different mixing
 characteristics.

\section{Privacy Preserving super-link and super-spreader detection}
\label{sec:ppgraph}

Proximity information is confidential since it can reveal information
about individual habits, movement patterns, personal and business
relationships. Sars-cov-2 is a pandemic and its reach is
global. Naturally, this calls for a globally coordinated response to
the epidemic. Health agencies across the globe have collected
proximity traces via a variety of centralised to fully decentralised
collection and processing mechanisms. However, as transportation links
are restored, it becomes important for health agencies to
collaboratively mine their data in order to detect super-links and
super-spreaders.

At the same time, organisations and individuals are understandably
reluctant to share information about proximity encounters they collect
with others, presenting a barrier to deploying graph analysis
techniques. Aggregating all the data in a centralised store for
analysis would be inappropriate, unsafe, as well as attract legal
challenges. Even though medical privacy has been suspended in some
countries such as the UK~\cite{gov2020privacy}, privacy concerns
dominate the use of proximity data for fighting epidemics.

In this section, we present privacy-preserving algorithms for
performing the computations necessary for locating super-spreaders and
super-links over a distributed proximity graph.  These algorithms
support inter-organisational collaboration without assuming the
availability of a centralised service or trusting the collaborating
health agencies with the confidentiality of each other's data. There are
three steps involved here: assembling a distributed graph,
pseudonymous identities, and distributed execution of random walks.

We assume that the agencies will behave in an adversarial manner, and
specifically as per a semi-honest model --- whilst they comply with
the security protocol they may carry out passive analysis of other
agencies data.

\subsection{Assembly and anonymisation of the graph}

\paragraph{Compute Intersection} The proximity information is
represented as a graph $G = (V,E)$, where the vertices are individuals
and edges are proximity encounters. The graph is distributed with each
participating agency operating over its part of the graph. $G =
\bigcup_{i=1}^{m} G_i$ where $G_i= (V_i, E_i)$ is the subgraph
corresponding to the $i^{th}$ health agency. The first step is to set
up an agreed space of numerical labels assigned to each susceptible,
infectious, infected, or recovered individual that serves the function
of addressing individuals. We achieve this by mapping labels to
individuals $L: \mathbb{Z}_{|V|} \rightarrow V$.

Before we split and assign label spaces to different agencies, we need
to find the gateway points, i.e individuals that appear on the graphs
of multiple agencies must be first identified. This is achieved via
private set intersection protocols to compute $V_i \cap V_j$, for all
$1 \leq i<j \leq m$.  A private set intersection protocol allows
multiple agencies to compute the intersection over their vertex set
such that each agency learns only the list of elements that is part of
the intersection.  A number of protocols for set intersection have
been proposed most of the efficient
designs~\cite{camenisch-zaverucha:fc09,freedman+:eurocrypt04,hazay-lindell:tcc08,jarecki-liu:tcc09,kissner-song:crypto05,deCristofaro-tsudik:eprint}
are proposed in the semi-honest model. Notable among these is the
Cristofaro-Tsudik scheme where compute time scales linearly in the
number of elements. Some progress has also been made in developing
schemes that work with weakly-malicious
adversaries~\cite{rindal2017ccs} but these are up to three times
slower. We propose the use of a protocol that has a \emph{linear
  runtime cost} and \emph{linear communication cost} in the number of
elements in the two sets; i.e., both runtime and communication costs
are $O(|V_i|+|V_j|)$~\cite{pinkas2018acmtsp}.  This makes set
intersection practical even for quite large sets consisting of graphs
of billions of nodes.

\paragraph{Anonymous label assignment} After common individuals are
identified, the label assignment proceeds as follows: the first health
agency assigns a random label in the range $0 \dots |V_1|-1$ to each
individual in $V_1$.  It also sends to every other health agency $j$
the labels of individuals in the intersection $V_1 \cap V_j$. The next
health agency assigns random labels in its range $|V_1|..|V_1 \cup
V_2|-1$ to the individuals within its graph $V_2 \setminus V_1$.  It
then sends to each health agency $j > 2$ the indices of common
individuals, i.e., those in the intersection $(V_2 \setminus V_1) \cap
V_j$. This process is repeated for each agency; at the end, all
individuals in $V$ have a unique label assigned to them, and each
agency knows the label for each individual within its graph $V_i$.

\paragraph{Assembling the distributed graph}
To perform a random walk, each health agency needs to learn the degree
of individual nodes, defined as  $d(v) = \sum_{i=1}^{m}
d_i(v)$, where $d_i(v)$ is the degree of individual $v$ in $G_i$.  The
sum can be computed by a simple protocol~\cite{nagaraja2010botgrep},
which is an extension of Chaum's dining cryptographer's
protocol~\cite{chaum:jcrypto88}.  Each agency $i$ creates $m$ random
shares $s^{(i)}_j \in \mathbb{Z}_l$ such that $\sum_{j=1}^{m}
s^{(i)}_j \equiv d_i(v) \bmod{l}$ (where $l$ is chosen such that $l >
\max_v d(v)$).  Each share $s^{(i)}_j$ is sent to agency $j$.  After
all shares have been distributed, each agency computes $s_i =
\sum_{j=1}^{m} s^{(j)}_i \bmod{l}$ and broadcasts it to all the other
agencies.  Then $d(v) = \sum_{i=1}^{m} s_i \bmod{l}$.  This protocol
is information-theoretically secure: any set of malicious agencies $S$
only learns the value $d(v) - \sum_{j \in S} d_i(v)$.  The protocol
can be executed in parallel for all individuals $v$ to learn all
individual degrees. Note that each agency must include only those
edges that cover individuals seen by both agencies, i.e., $e =
(v_1,v_2)$ such that $v_1,v_2 \in V_i \cap V_j$.  Whenever a duplicate
edge is detected, one of the agencies drops the edge from its set of
edges $E_i$.

\subsection{Distributed disease spread analysis}
Together, the agencies analyse the disease spreading over the
distributed proximity graph by performing a secure multiparty
matrix-vector multiplication operation. This is constructed from well
understood primitives. Given a transition matrix $T$ (representation
of the proximity graph) and a distributed vector of initial infection
probabilities $\vec{v}$, we can compute $T\vec{v}$, the infection
spread vector after a single epoch represented by a one random walk
step as follows. Each agency creates matrices $T_i$ such that
$\sum_{i=1}^{m} T_i = T$. At the highest level of distribution, each
agency owns just one row, which defaults to individuals participating
directly with their proximity information.

Next, they compute $T_i \vec{v}$ in a distributed fashion and compute
the final sum at the end. To construct $T_i$, each agency sets the
transition matrix $(T_i)_{j,k}$ to be $1/deg(v_j)$ for each edge
$(j,k) \in E_i$ (after duplicate edges have been removed). As with
most large graphs the transition matrix is sparse and therefore incurs
a storage cost of $O(|E_i|) \ll O(|V_i|^2)$ instead of $O(N^2)$.

Paillier encryption~\cite{paillier:eurocrypt99} is used to perform
computation on an encrypted vector $E(\vec{v})$.  Paillier supports a
homomorphism that allows one to compute $E(x) \oplus E(y) = E(x+y)$;
it also allows the multiplication by a constant: $c \otimes E(x) =
E(cx)$.  This, given an encrypted vector $E(\vec{v})$ and a known
matrix $T_i$, it is possible to compute $E(T_i \vec{v})$.

Damg{\aa}rd and Jurik~\cite{damgard-jurik:pkc01} developed an
efficient distributed key generation mechanism for Paillier that
allows the creation of a public key $K$ such that no individual health
agency knows the private key, but together, they can decrypt the
value.  In the full protocol, one agency creates an encrypted vector
$E(\vec{v})$ that represents the initial state of the random walk.
This vector is then sent to each agency, who computes $E(T_i
\vec{v})$.  Finally, the agencies sum up the individual results to
obtain $E(\sum_{i=1}^{m} T_i \vec{v}) = E(T \vec{v})$. This process
can be iterated to obtain $E(T^k \vec{v})$.  Finally, the agencies
jointly decrypt the result to obtain $T^k \vec{v}$.

Note that Paillier operates over members $\mathbb{Z}_n$, where $n$ is
the product of two large primes.  However, the vector $\vec{v}$ and
the transition matrices $T_i$ are floating point numbers. To perform
real arithmetic over a finite field, we represent floating point
numbers as fixed point numbers by multiplying them by a fixed base
storing $\lfloor x \times 2^c \rfloor$ (equivalently, $(x-\epsilon)
\times 2^c$, where $\epsilon < 2^{-c}$).  Each multiplication results
in changing the position of the fixed point, since:

\[ \left((x-\epsilon_1) \times 2^c\right)\left((y-\epsilon_2) \times 2^c\right) = (xy-\epsilon_3) \times 2^{2c} \]

\noindent where $\epsilon_3 < 2^{-c+1}$.  Therefore, we must ensure
that $2^{kc} < n$, where $k$ is the number of random walk steps. The
maximal length random walk we use is $ (\log_{\bar{d}} |V|)^2$, where
$\bar{d}$ is the average node degree, so $k < 40$ (allowing epidemic
attack rates of $R_0 \leq 40$ ), which offers sufficient fixed-point
precision to work with for a typical choice of $n$ (1024 or 2048
bits).

\paragraphb{Security:} Note that each party transmits values encrypted
under the public key, and no party involved in this protocol has a
copy of the distributed private key.  The Paillier encryption scheme
is known to be CPA-secure~\cite{paillier:eurocrypt99}; thus these
protocols reveal no information about the vector or matrix in the
honest-but-curious setting.

\paragraph{Performance:} The base privacy-preserving protocols we propose
are efficient. However, when the number of individuals or agencies
increase, the operations can expect to take a significant amount of
processing time.  We estimate the actual processing costs and
bandwidth overhead, using some approximate parameters.  Specifically,
with a distributed graph of 10 million individuals, with an average
degree of 10 per node.\footnote{The average degree in proximity graphs
  is likely to be smaller than this value but we are seek conservative
  estimates.} Each step of the spreading process would require
$O(|E|)$ multiplications and additions over encrypted values. An
efficient implementation~\cite{sfsdist,jost2015cryptoeprint} over an
i7 CPU with 8GB RAM can compute 7200 multiplications and approx.~1
million additions per second using a 2048-bit encryption key. Thus a
single step of disease spreading would take approx.~$20\,000$ seconds
to compute.  We note that it would be straightforward to apply
parallelisation techniques to reduce this compute time by an order of
magnitude. We have excluded the costs of setup time which is a one-off
effort.

\section{Experimental Results}

To evaluate the effectiveness of intervention strategies, we simulate
disease spreading over a number of real-world graphs and study the
impact of interventions.

The simulation is divided into two phases:\\
In the {\em intervention phase}, the responder takes action according
to one or more {\em intervention strategies} -- the selection of
network elements (nodes or edges) subjected to restrictions
constitutes an intervention strategy. Restrictions on nodes may
physically manifest in the form of quarantining, self-isolation, or
physical distancing measures over individuals. We consider the
following strategies in our study: {\em null response} -- where the
responder does nothing, {\em contact tracing} -- a widely used
strategy in the wake of the Covid-19 pandemic, {\em super-spreader}
suppression, and {\em super-link} suppression. The intervention phase
has three associated parameters: first, the intervention strategy
chosen by the public-health responder; second {\em visibility}, the
proportion of the propagation graph that is visible to the responder;
and third {\em isolation budget}, is the proportion of the propagation
graph that is subjected to restrictions. 

The {\em spreading phase} consists of several rounds. In the first
round, an initial infection source is selected by choosing a node
uniformly at random within the network. Subsequently, a randomly
chosen neighbour of the first infected individual that is not in
isolation contracts the infection. In subsequent rounds, the process
repeats, and in each round for every infected individual that has not
been isolated, a randomly chosen neighbour who is not isolated,
contracts the infection. This process repeats until there are no more
susceptible individuals. In each round, we measure the transmission
potential to measure the extent of disease spread within the
network. It is worth noting that the full propagation network is
available to the pathogen for disease spreading, i.e only the
responder has partial visibility whereas the disease spreads on the
full topology.


\paragraphb{Dataset:} We use real world contact traces from the
Brightkite social-proximity network. This consists of $58228$ nodes and $214078$
edges. An edge between two nodes indicates a social link. In
Brightkite, an edge between individuals is created when they interact
with each other at least ten times. Studies find a close link between
the existence of social links and proximity
encounters~\cite{cho2011kdd}, specifically a substantial social link
greatly increases the chances of a proximity encounter. Specifically,
we know that human proximity encounters exhibit structural patterns
owing to the influence of social relationships and geographical
constraints. We therefore selected a location-based online social
networks to study the dynamics of disease spreading in response to
public-health interventions.

\subsection{Baseline -- No response}
\begin{figure*}[!ht]
\centering 
\includegraphics[width=2.0in]{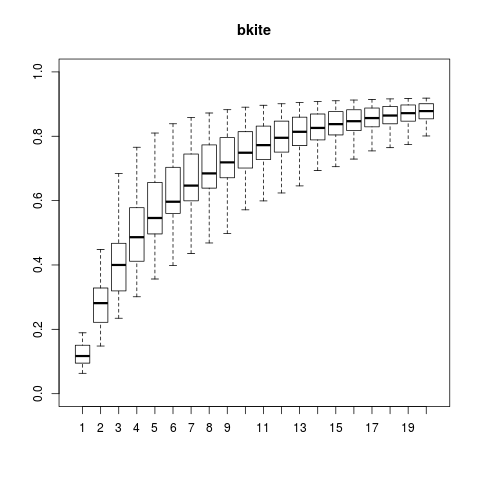}
 \caption{With no interventions infections spread rapidly achieving
   maximal spread within a few rounds.}
  \label{fig:noresponse}
\end{figure*}

Initially, we simulated disease spreading when no interventions are
undertaken within the intervention phase, this gives us the {\em base
  case}. Figure~\ref{fig:noresponse} shows disease propagation over
time. Time is divided into multiple rounds with each round consisting
of a unit of time over which all proximity encounters involving a given
node and its neighbours takes place. As explained in
section~\ref{metrics}, we use the transmission potential metric --
defined as the the normalised entropy of infection probabilities of
all individuals within a population -- to measure disease
propagation. In the absence of any isolation, the transmission
potential reaches maximal levels within three rounds. This is
consistent with the exponential increase in the number of cases
observed in other epidemiological
studies~\cite{ferretti2020Science,guo2019BMC}.

\subsection{Contact tracing and testing}
The most popular non-pharmaceutical intervention strategy is contact
tracing~\cite{culnane2020uk,bay2020bluetrace,gov2020privacy,ferretti2020Science,eichner2003Oxford,wiertz2020predicted,cdc2020principles,yoneki2012flu,dp3t},
so it is natural to use this as a second baseline of comparison. In
contact tracing, the intervention phase alternates with the spreading
phase. The first round starts with a fraction of individuals (1\% of
the population) being randomly chosen as initial sites of
infection. In the spreading phase, each of the chosen initial
infection sources infects its neighbours. This is followed by the
intervention phase wherein all infected nodes, and their neighbours
(i.e infected individuals and those in their proximity regardless of
infection state) are isolated. The goal of contact tracing is to stop
disease spread by isolating clusters of infected
individuals. Following the first round, in each subsequent round, the
intervention phase is followed by the spreading phase. We simulated
contact tracing for a total of 20 rounds.

\begin{figure*}[!ht]
  \centering

    \subfigure[V  = 10\%, Q=05\%]{ \label{fig:ct:a1}
    {\includegraphics[width=1.2in]{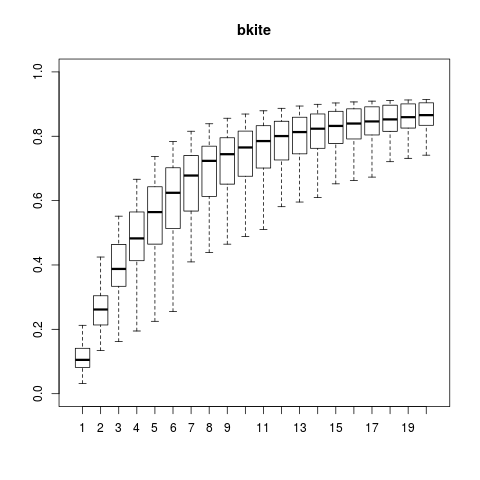}}}
  \subfigure[V  = 10\%, Q=15\%]{ \label{fig:ct:a2}
       {\includegraphics[width=1.2in]{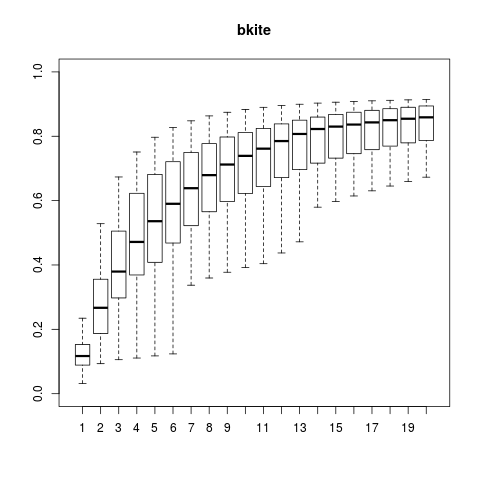}}}
  \subfigure[V  = 10\%, Q=25\%]{ \label{fig:ct:a3}
       {\includegraphics[width=1.2in]{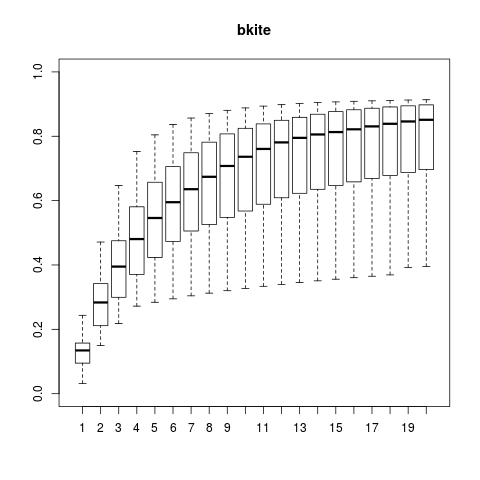}}}
  \subfigure[V  = 10\%, Q=35\%]{ \label{fig:ct:a4}
       {\includegraphics[width=1.2in]{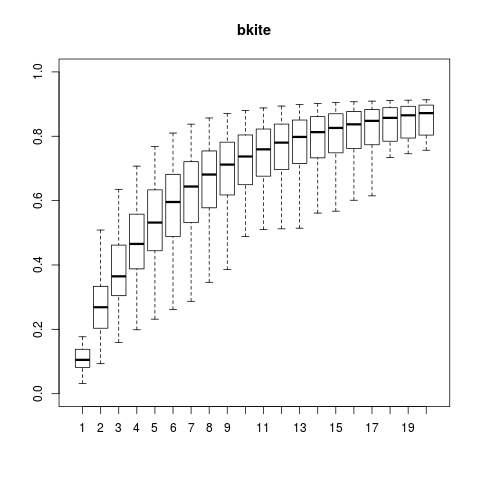}}}
  \subfigure[V  = 10\%, Q=45\%]{ \label{fig:ct:a5}
    {\includegraphics[width=1.2in]{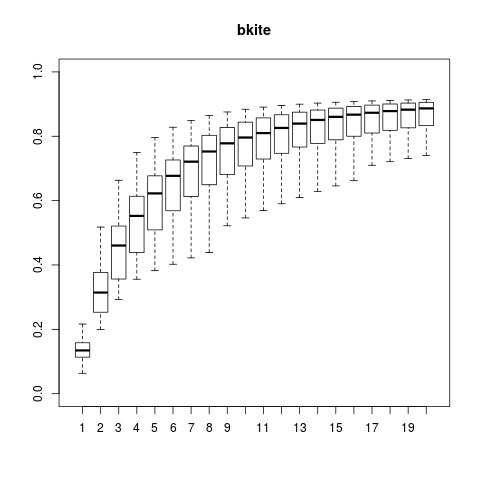}}}\\
  \subfigure[V  = 35\%, Q=05\%]{ \label{fig:ct:b1}
       {\includegraphics[width=1.2in]{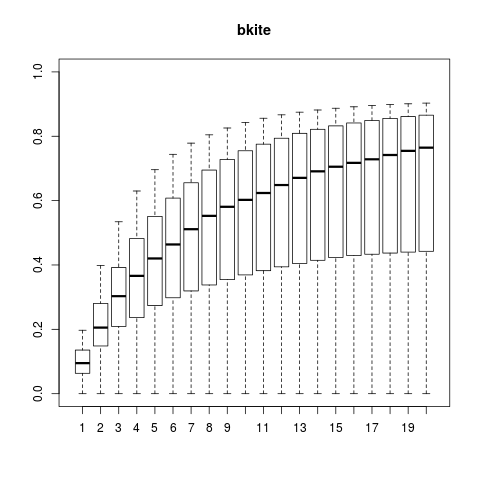}}}
  \subfigure[V  = 35\%, Q=15\%]{ \label{fig:ct:b2}
       {\includegraphics[width=1.2in]{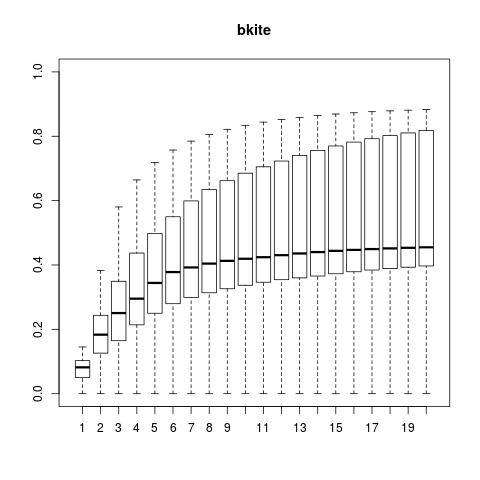}}}
  \subfigure[V  = 35\%, Q=25\%]{ \label{fig:ct:b3}
       {\includegraphics[width=1.2in]{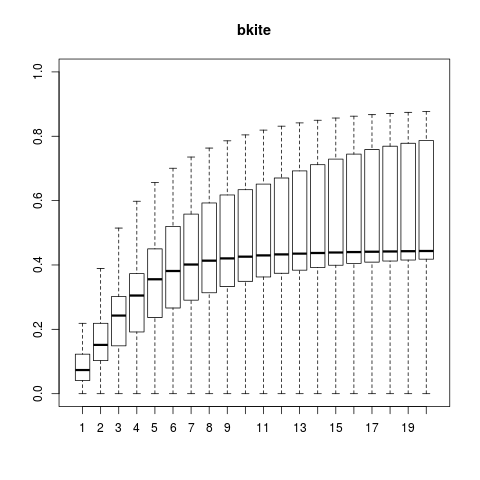}}}
  \subfigure[V  = 35\%, Q=35\%]{ \label{fig:ct:b4}
       {\includegraphics[width=1.2in]{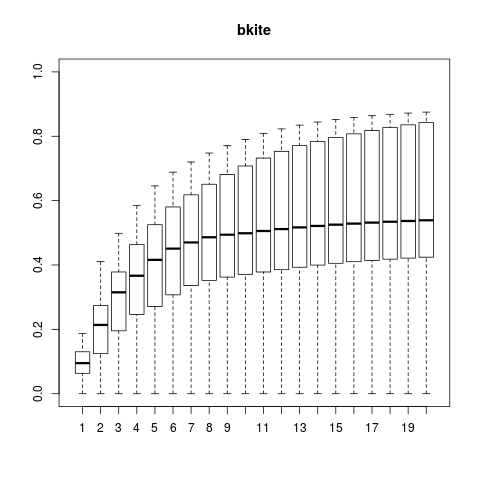}}}
  \subfigure[V  = 35\%, Q=45\%]{ \label{fig:ct:b5}
    {\includegraphics[width=1.2in]{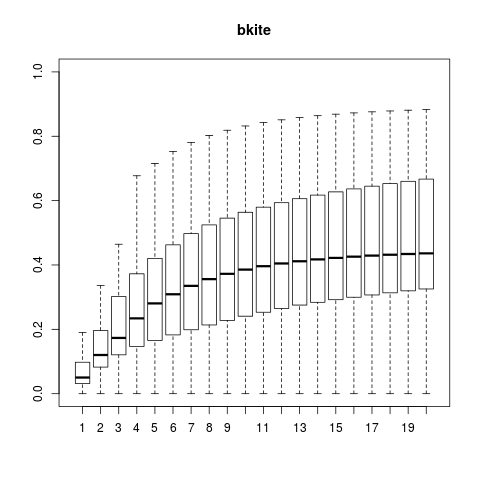}}}\\
  \subfigure[V  = 60\%, Q=05\%]{ \label{fig:ct:c1}
       {\includegraphics[width=1.2in]{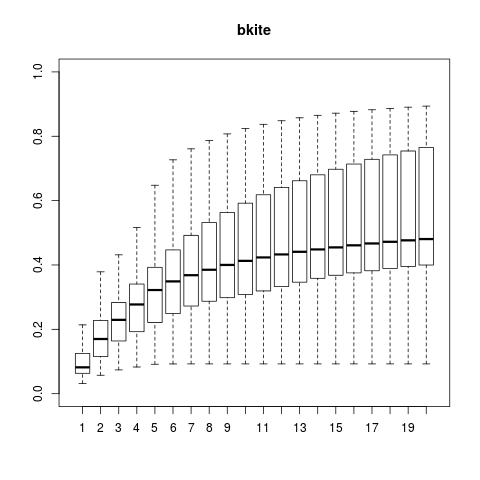}}}
  \subfigure[V  = 60\%, Q=15\%]{ \label{fig:ct:c2}
       {\includegraphics[width=1.2in]{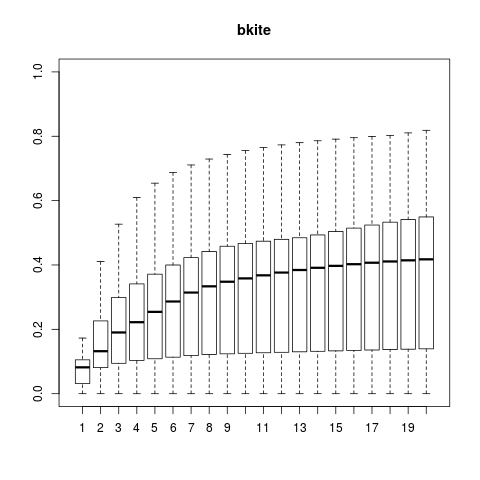}}}
  \subfigure[V  = 60\%, Q=25\%]{ \label{fig:ct:c3}
       {\includegraphics[width=1.2in]{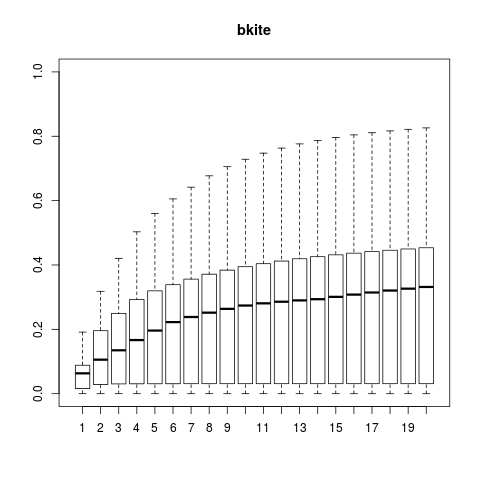}}}
  \subfigure[V  = 60\%, Q=35\%]{ \label{fig:ct:c4}
       {\includegraphics[width=1.2in]{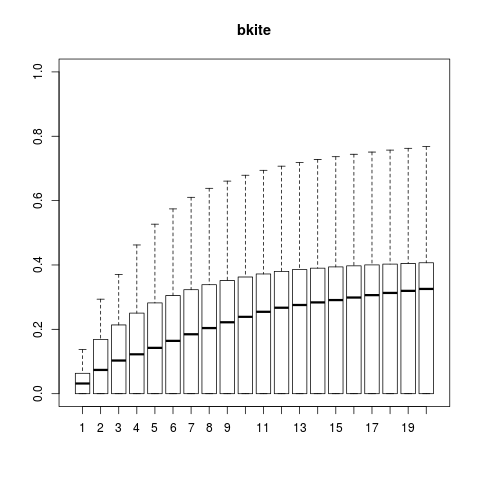}}}
  \subfigure[V  = 60\%, Q=45\%]{ \label{fig:ct:c5}
    {\includegraphics[width=1.2in]{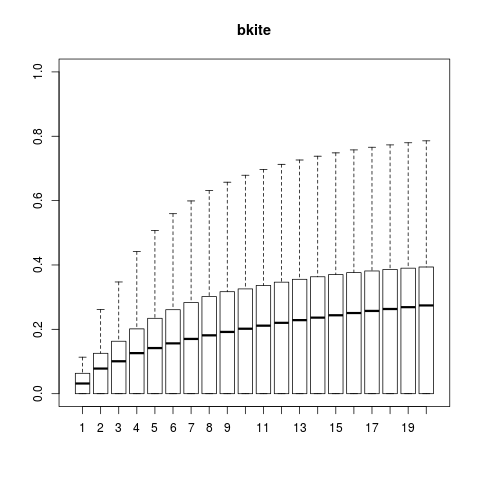}}}\\
  \subfigure[V  = 85\%, Q=05\%]{ \label{fig:ct:d1}
       {\includegraphics[width=1.2in]{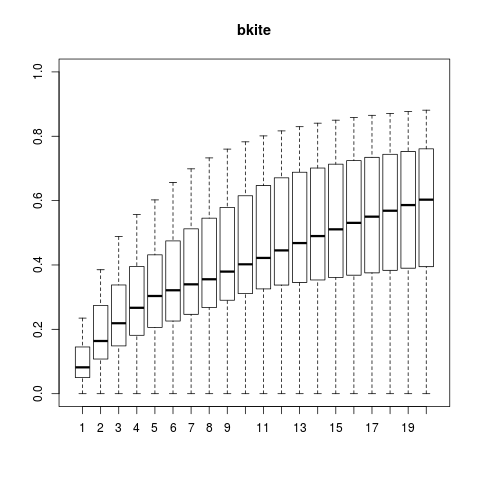}}}
  \subfigure[V  = 85\%, Q=15\%]{ \label{fig:ct:d2}
       {\includegraphics[width=1.2in]{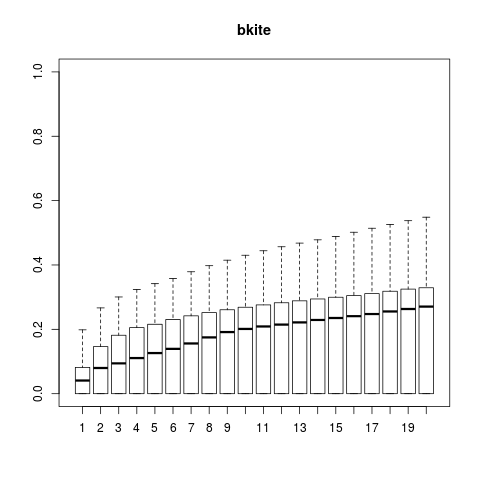}}}
  \subfigure[V  = 85\%, Q=25\%]{ \label{fig:ct:d3}
       {\includegraphics[width=1.2in]{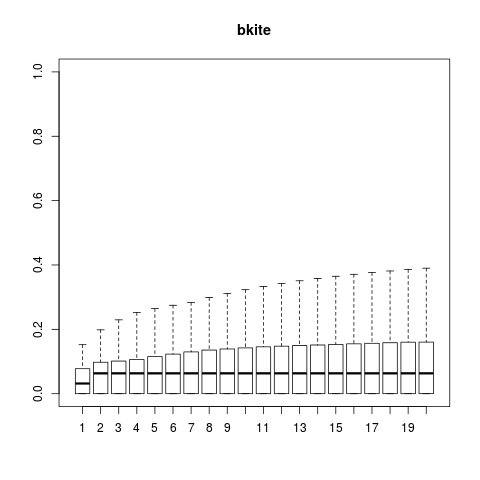}}}
  \subfigure[V  = 85\%, Q=35\%]{ \label{fig:ct:d4}
       {\includegraphics[width=1.2in]{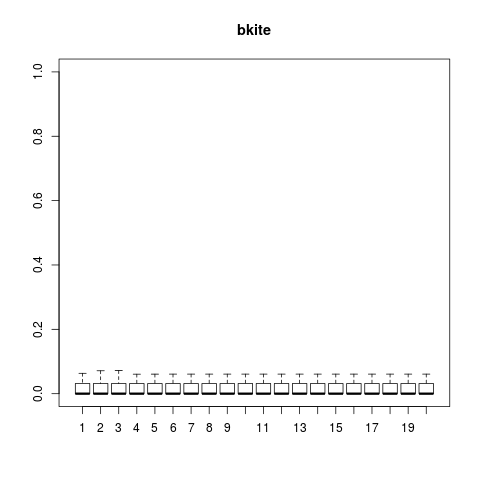}}}
  \subfigure[V  = 85\%, Q=45\%]{ \label{fig:ct:d5}
    {\includegraphics[width=1.2in]{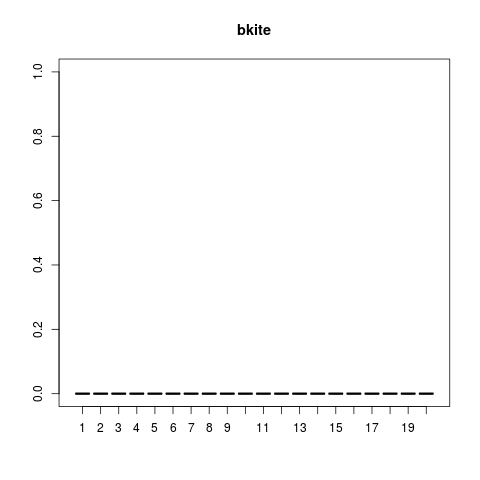}}}%
  \caption{Effectiveness of Contact Tracing and Testing in the
    Brightkite dataset. $V$ is the fraction of the contact graph
    visible to the responder. $Q$ is the fraction of the population
    asked to self-isolate. For low visibility (corresponding to low
    adoption rates of contact tracing apps), there is little impact on
    spreading even with an isolation budget of up to 50\% of the
    population. Above 60\% adoption, isolating even 15\% of the
    population slows down the spread with substantial reductions
    achieved at 50\%. At 85\% adoption rate, merely isolating 35\% of
    infected people and their proximity neighbours is enough to stop
    disease spread.}
  \label{fig:ct}
\end{figure*}

\setcounter{subfigure}{0}

As figure~\ref{fig:ct} shows, testing and tracing is certainly better
than a null intervention.  When all or most of the propagation graph
is visible, contact tracing works exceedingly well and stops disease
propagation in its entirety. This is of course subject to an adequate
isolation budget. When 85\% of the population has installed contact
tracing apps i.e visibility of propagation network is equal to or
greater than 85\% and the responder can ask at least 45\% of the
population to be placed in isolation, the disease spread is
stopped. Note that the virus itself isn't eliminated, and an outbreak
can quickly take hold within three rounds, should contact testing and
tracing stop. Note that this is the same result as full lockdown at
half the isolation budget (and social cost).

Sadly contact tracing does not fare well at low rates of trace
visibility. When less than 60\% of the population has installed
contact tracing, we observe a significant increase in transmission
potential even when half the population is under isolation. And, at
10\% visibility, contact tracing is almost as ineffective as the null
response. We note however that even at partial visibility of 35\%, the
transmission potential is halved compared to the null intervention baseline,
with 15\% isolation budget.

When less than 85\% of the propagation network is visible, we observe
that an increase in the isolation budget has no corresponding impact
on the disease spreading potential of the network. This is because the
disease spreading process is occurring outside the visible subset of
the propagation network even at medium to high rates of visibility. 

Overall, simulations shows that contact tracing is effective at
reliably halving transmission potential when at least 60\% of the
propagation network is visible to the responder. These tests are
averaged over 10000 trials.

\subsection{Isolating super-links}

Having established that contact tracing only works only after a
substantial fraction of the population is visible, we now explore the
graph theoretic approach of super-link detection. As before, we start
with seed individuals that initially contract the infection. This
group of initial infectees is 1\% of the population i.e $580$ randomly
chosen individuals. We then simulate disease spreading from one person
to another under the influence of restrictions at various rates of
network visibility and isolation budgets.

Disease spreading is studied as a multi-step event. The intervention
strategy is applied by the responder using partial knowledge of network
topology and no information about the site of initial
infections. Therefore, a trivial response that involves quarantining
all the initial infected population is excluded, as it is not realistic
to frequently test all individuals over short periodic cycles.

In the initial infection step, a randomly chosen subset of the
population is declared infected. These individuals constitute the {\em
  initial infectious group}. Next, super-links (massively influential
graph edges) are identified by the responder who has no knowledge of
the initial infectious group. Individuals constituting a super-link
graph edge are subject to isolation limited by the isolation budget
i.e a maximum of $x|V|$ nodes may be asked to isolate themselves,
where $x$ is a fraction and $|V|$ is the total population size. On the
graph, node isolation is implemented by removing all its edges from
the graph topology. These nodes form the set of infected but
non-infectious nodes. Since the isolation budget is spent upfront,
there is no isolation capacity available for demand arising from
secondary infections. Thus disease transmission from infectious
individuals arising out of any successful secondary infections will
only be stopped by the removal of all disease transmission paths
between infectious and susceptible individuals. This is a significant
ask of any public health response.

In subsequent steps, disease propagation is simulated as spreading
from one individual to another. We assume that all proximity
encounters between an infectious and a susceptible individual results
in an infection (i.e a 100\% attack rate). $R_0$ is $3.68$ in the
Brightkite social network. All $R_0>1$ are relevant outbreak
scenarios, since these infection rates will lead to an outbreak.

\begin{figure*}
  \centering

  \subfigure[V  = 10\%, Q=05\%]{ \label{fig:wt:a1}
       {\includegraphics[width=1.2in]{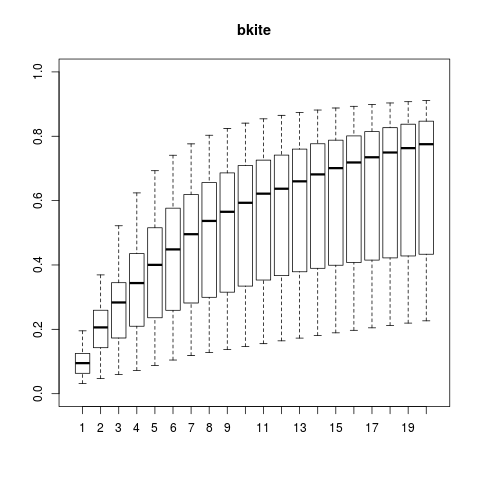}}}
  \subfigure[V  = 10\%, Q=15\%]{ \label{fig:wt:a2}
       {\includegraphics[width=1.2in]{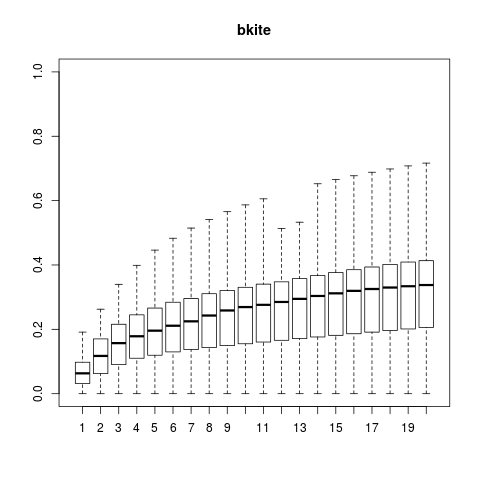}}}
  \subfigure[V  = 10\%, Q=25\%]{ \label{fig:wt:a3}
       {\includegraphics[width=1.2in]{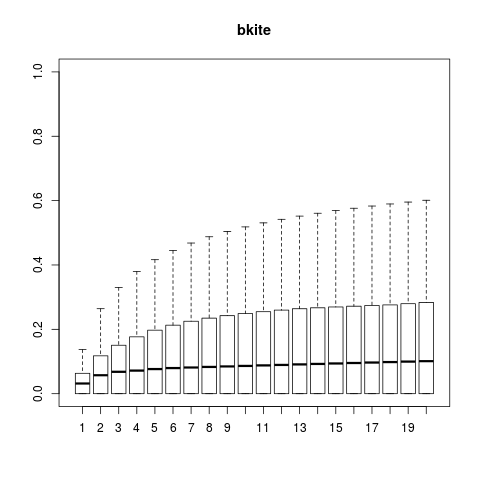}}}
  \subfigure[V  = 10\%, Q=35\%]{ \label{fig:wt:a4}
       {\includegraphics[width=1.2in]{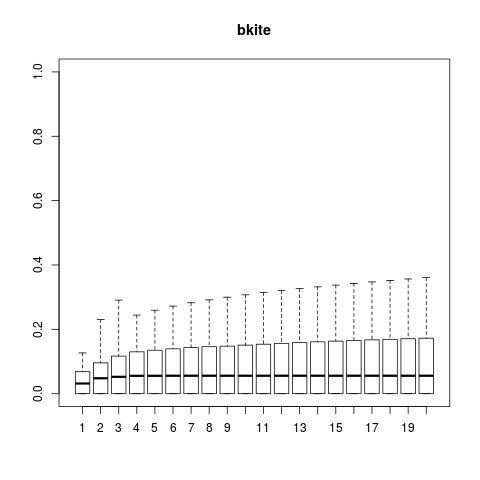}}}
  \subfigure[V  = 10\%, Q=45\%]{ \label{fig:wt:a5}
    {\includegraphics[width=1.2in]{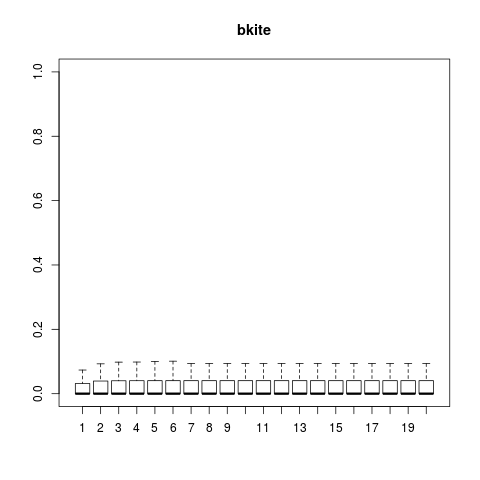}}}\\
  \subfigure[V  = 35\%, Q=05\%]{ \label{fig:wt:b1}
       {\includegraphics[width=1.2in]{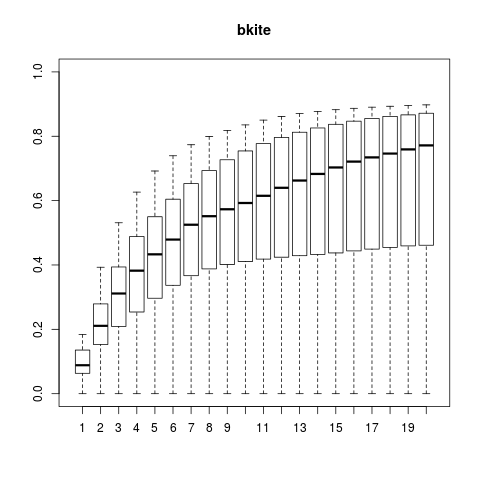}}}
  \subfigure[V  = 35\%, Q=15\%]{ \label{fig:wt:b2}
       {\includegraphics[width=1.2in]{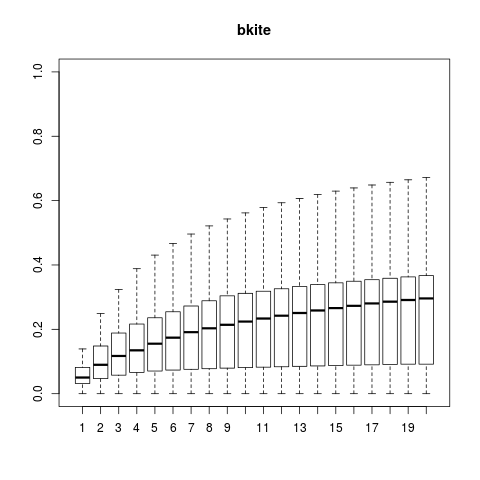}}}
  \subfigure[V  = 35\%, Q=25\%]{ \label{fig:wt:b3}
       {\includegraphics[width=1.2in]{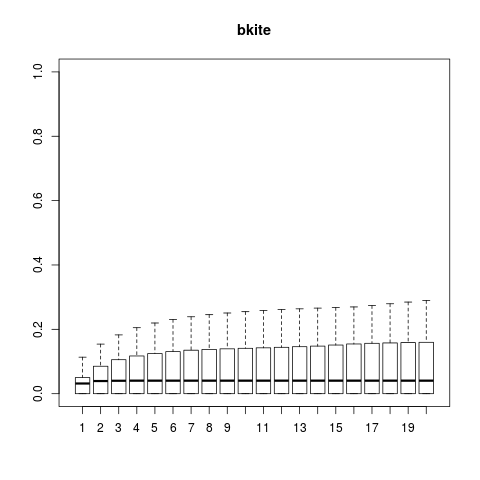}}}
  \subfigure[V  = 35\%, Q=35\%]{ \label{fig:wt:b4}
       {\includegraphics[width=1.2in]{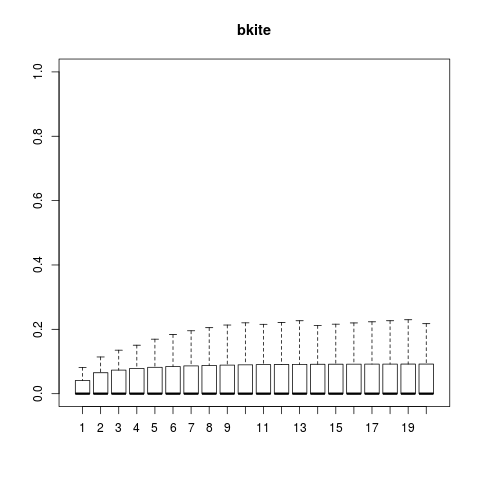}}}
  \subfigure[V  = 35\%, Q=45\%]{ \label{fig:wt:b5}
    {\includegraphics[width=1.2in]{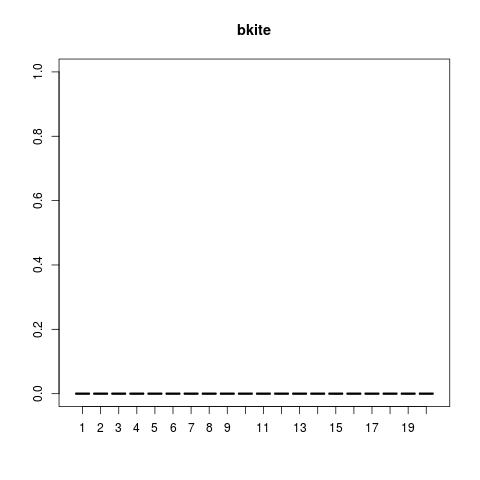}}}\\
  \subfigure[V  = 60\%, Q=05\%]{ \label{fig:wt:c1}
       {\includegraphics[width=1.2in]{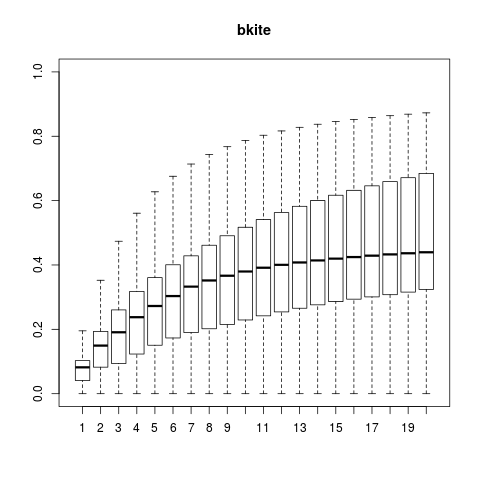}}}
  \subfigure[V  = 60\%, Q=15\%]{ \label{fig:wt:c2}
       {\includegraphics[width=1.2in]{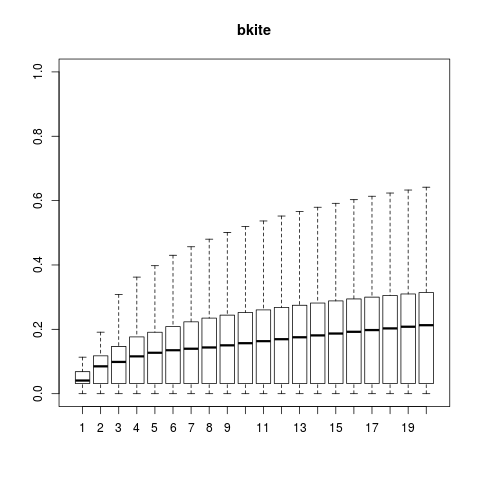}}}
  \subfigure[V  = 60\%, Q=25\%]{ \label{fig:wt:c3}
       {\includegraphics[width=1.2in]{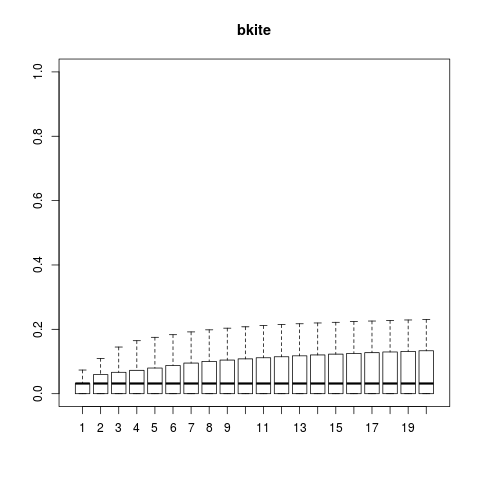}}}
  \subfigure[V  = 60\%, Q=35\%]{ \label{fig:wt:c4}
       {\includegraphics[width=1.2in]{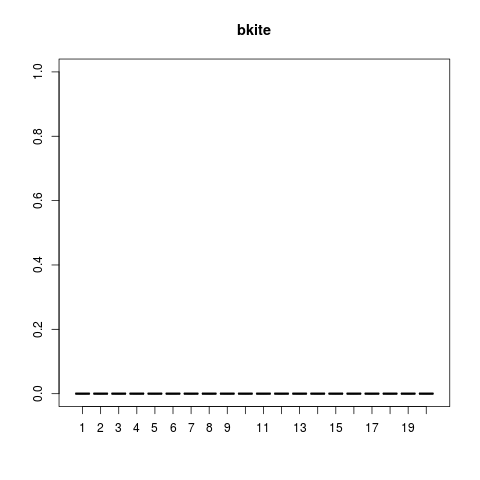}}}
  \subfigure[V  = 60\%, Q=45\%]{ \label{fig:wt:c5}
    {\includegraphics[width=1.2in]{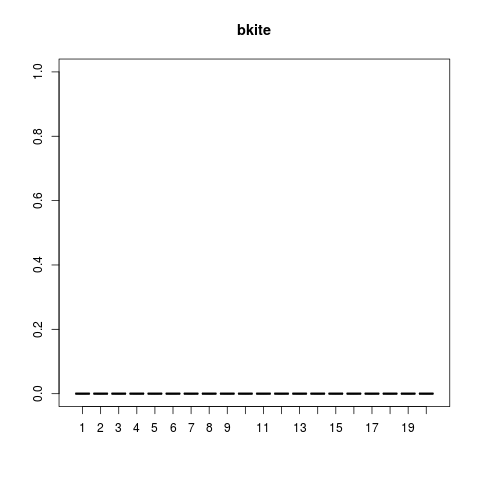}}}\\
  \subfigure[V  = 85\%, Q=05\%]{ \label{fig:wt:d1}
       {\includegraphics[width=1.2in]{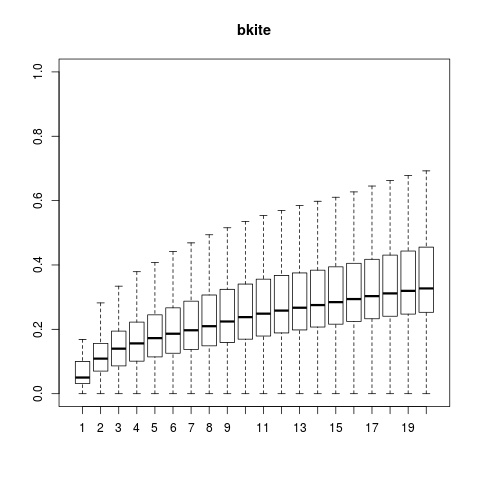}}}
  \subfigure[V  = 85\%, Q=15\%]{ \label{fig:wt:d2}
       {\includegraphics[width=1.2in]{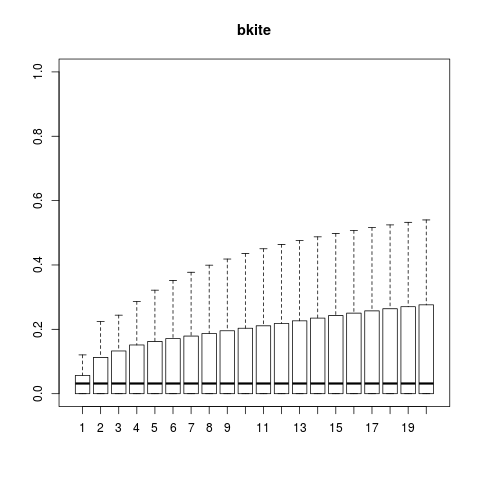}}}
  \subfigure[V  = 85\%, Q=25\%]{ \label{fig:wt:d3}
       {\includegraphics[width=1.2in]{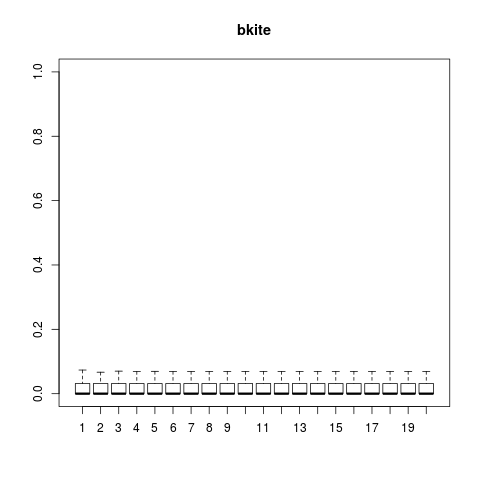}}}
  \subfigure[V  = 85\%, Q=35\%]{ \label{fig:wt:d4}
       {\includegraphics[width=1.2in]{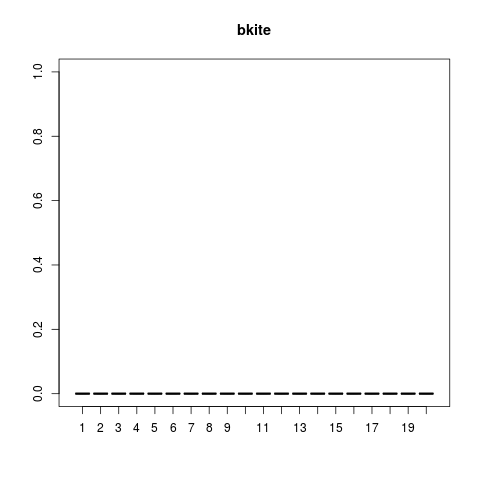}}}
  \subfigure[V  = 85\%, Q=45\%]{ \label{fig:wt:d5}
    {\includegraphics[width=1.2in]{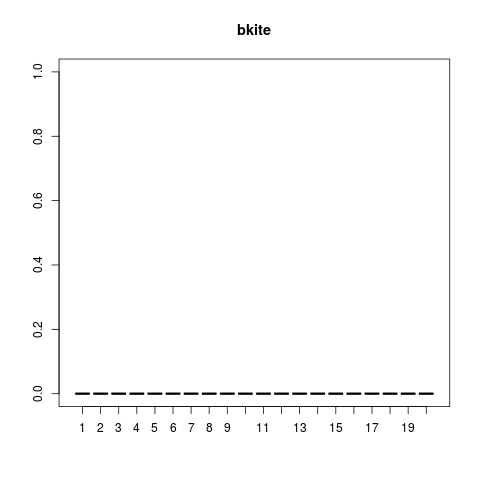}}}
  \caption{Effectiveness of super-link intervention in Brightkite
    graph. At just 10\% visibility, the disease spreading can be fully
    stopped by isolating 45\% with significant reductions achieved at
    a isolation budget of 15\%. At medium visibility of 35\%, disease
    suppression is achieved by isolating 45\% of the population. At
    high visibility of 85\%, just 25\% of the population needs to be
    isolated to fully suppress the disease with significant reduction
    achieved at 15\% isolation.  $V$ is the fraction of the contact
    graph visible to the responder. $Q$ is the fraction of the
    population placed under self-isolation.}
  \label{fig:wt}
\end{figure*}
\setcounter{subfigure}{0}

Figure~\ref{fig:wt} shows amortised results over 10000 randomised
simulations. We can observe that proactive isolation of super-links can
be rather effective in limiting disease propagation. Both visibility
and isolation budgets play an important role. Overall we can observe
that a moderate visibility and moderate isolation budget result in
flattening intervention potential (disease propagation) by a factor of
5 to 8, while higher isolation helps reduce the disease spread to
small local clusters and entirely eliminating outbreaks.

\paragraph{Low visibility --- low isolation}
At low levels of graph visibility of 10\%, and with a small isolation
budget of 5\% (Figure~\ref{fig:wt:a1}), the effectiveness of
super-link intervention is expectedly limited.  We observe an
expansion of numeric range with a decreasing lower bound. This
suggests a decrease in intervention potential (likelihood of
contracting the disease) in some cases within each time
interval. However, the average remains roughly the same as in the case
where public-health responders don't intervene
(Figure~\ref{fig:noresponse}). And, by the 20th time interval, the
likelihood of contracting the disease is as high as the baseline case
of no response, although the rate of increase is markedly slower in
comparison with the baseline. At exceedingly low budgets super-link
detection slows down disease propagation but does not limit the
spread.

\paragraph{Low visibility --- medium to high isolation}
At a slightly increased isolation budget of 15\% with graph visibility
still limited to 10\%, super-link suppression results in a reduction
of disease spread by a decrease of 50\% compared to the baseline case
of no intervention. There is further improvement with increase in
budgets, with a further five-fold decrease in the likelihood of
contracting the disease when 45\% of the population is under
restrictions of physical isolation. In contrast, contact tracing only
exhibits a two fold decrease for equivalent visibility and isolation
budget.

\paragraph{Medium visibility --- low to high isolation}
As visibility grows to 35\%, with a medium budget of 25\% isolation we
observe a significant improvement compared to the baseline
(Figure~\ref{fig:wt:b3}) -- the intervention potential flattens with
disease spread limited to local clusters. A scenario achieved only at
high budgets (45\%) for low visibility (10\%). When combined with
higher isolation budgets (45\%) we observe that the disease spread is
reduced to its lowest theoretical limit (Figure-~\ref{fig:wt:b5}).

\paragraph{High visibility}
With higher visibility the curve flattens further and we observe that
each successive 25\% increase in visibility enables 10\% reduction in
isolation budgets to achieve the same outcome. For instance, 60\%
visibility and 35\% isolation (Figure~\ref{fig:wt:d4}) has similar
impact as 35\% visibility and 45\% isolation
(Figure~\ref{fig:wt:c5}). Similarly, 85\% visibility and 35\%
isolation (Figure~\ref{fig:wt:d4}) has similar impact as 60\%
visibility and 45\% isolation (Figure~\ref{fig:wt:c5}). In general,
the case of high visibility $\geq 60\%$ is essentially a theoretical
scenario as few responders have managed to gain high visibility into
the populations under their care using purely voluntary means.

Overall we find that as visibility grows it results in a better impact
on disease propagation with the same isolation budget, reducing the
intervention potential by half from 0.8 to 0.3 (stable values).

\subsection{Isolating Super-spreaders}

\begin{figure*}
  \centering
  \subfigure[V  = 10\%, Q=05\%]{ \label{fig:deg:a1}
       {\includegraphics[width=1.24in]{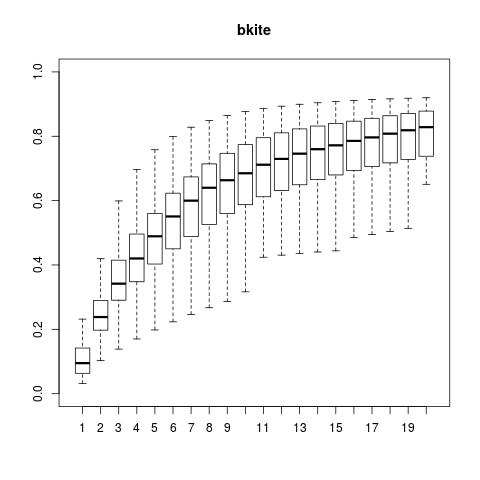}}}
  \subfigure[V  = 10\%, Q=15\%]{ \label{fig:deg:a2}
       {\includegraphics[width=1.24in]{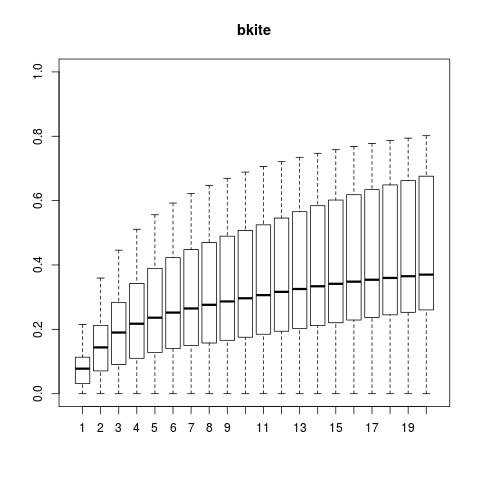}}}
  \subfigure[V  = 10\%, Q=25\%]{ \label{fig:deg:a3}
       {\includegraphics[width=1.24in]{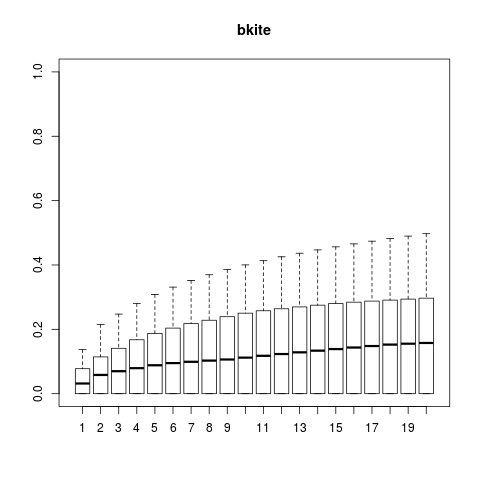}}}
  \subfigure[V  = 10\%, Q=35\%]{ \label{fig:deg:a4}
       {\includegraphics[width=1.24in]{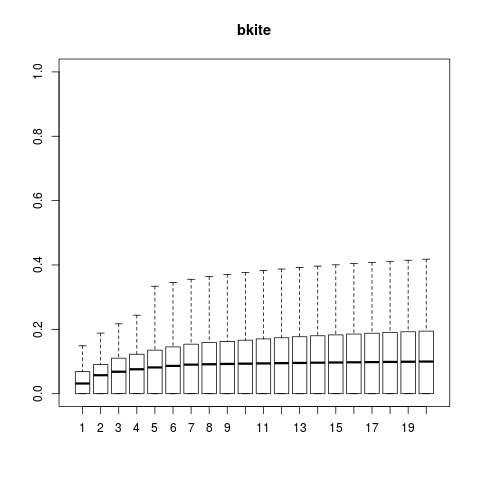}}}
  \subfigure[V  = 10\%, Q=45\%]{ \label{fig:deg:a5}
    {\includegraphics[width=1.24in]{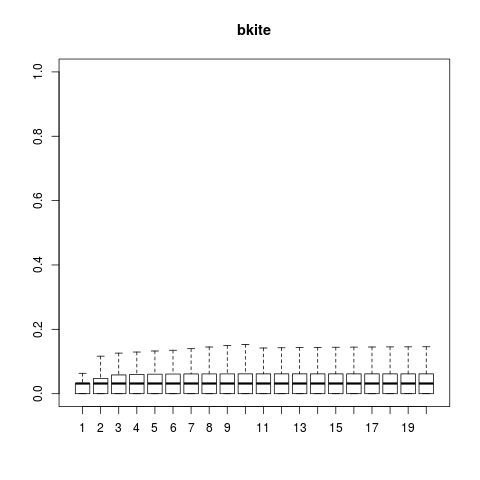}}}\\
  \subfigure[V  = 35\%, Q=05\%]{ \label{fig:deg:b1}
       {\includegraphics[width=1.24in]{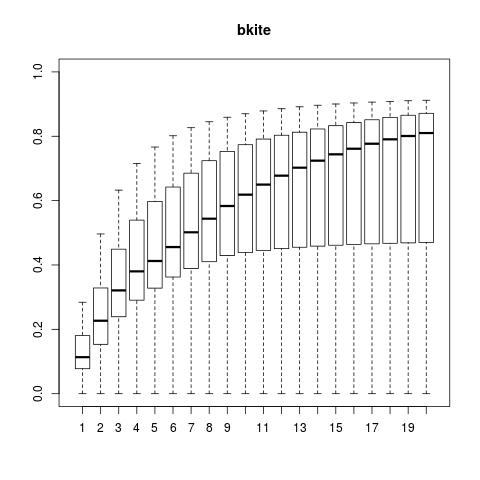}}}
  \subfigure[V  = 35\%, Q=15\%]{ \label{fig:deg:b2}
       {\includegraphics[width=1.24in]{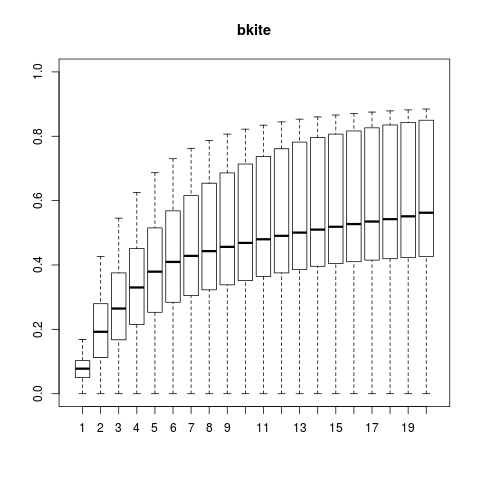}}}
  \subfigure[V  = 35\%, Q=25\%]{ \label{fig:deg:b3}
       {\includegraphics[width=1.24in]{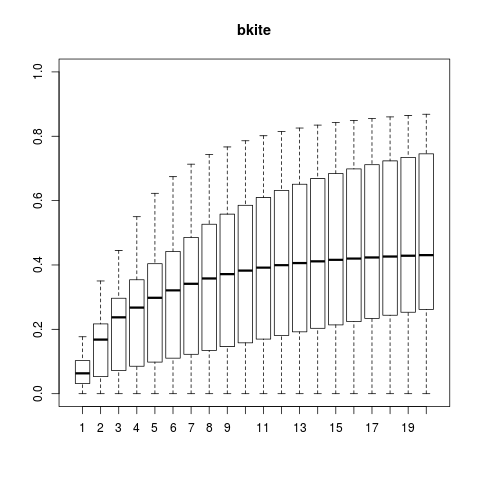}}}
  \subfigure[V  = 35\%, Q=35\%]{ \label{fig:deg:b4}
       {\includegraphics[width=1.24in]{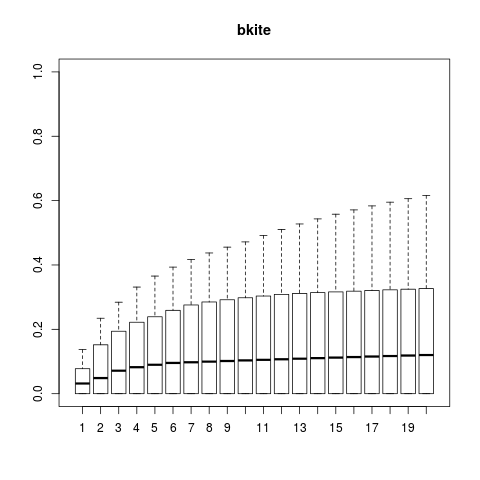}}}
  \subfigure[V  = 35\%, Q=45\%]{ \label{fig:deg:b5}
    {\includegraphics[width=1.24in]{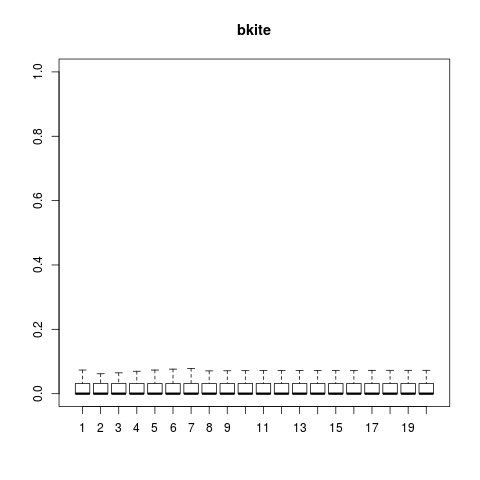}}}\\
  \subfigure[V  = 60\%, Q=05\%]{ \label{fig:deg:c1}
       {\includegraphics[width=1.24in]{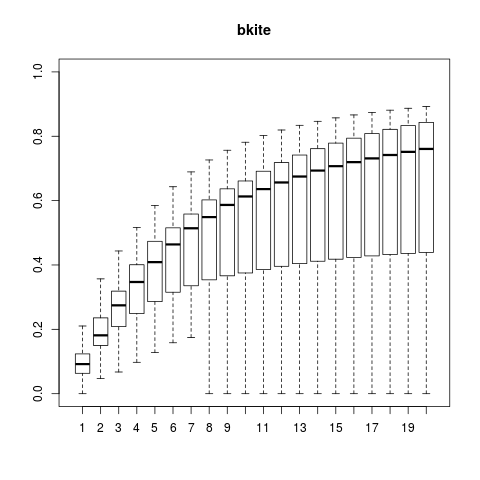}}}
  \subfigure[V  = 60\%, Q=15\%]{ \label{fig:deg:c2}
       {\includegraphics[width=1.24in]{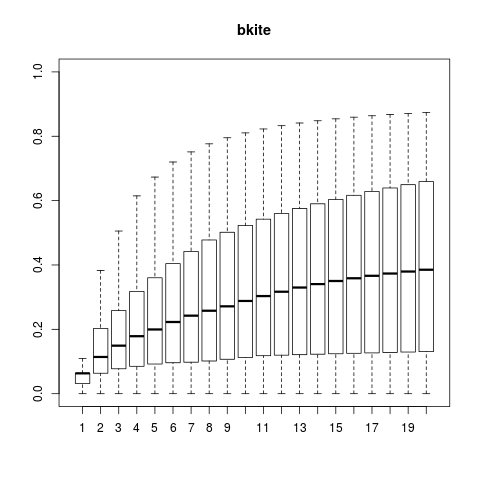}}}
  \subfigure[V  = 60\%, Q=25\%]{ \label{fig:deg:c3}
       {\includegraphics[width=1.24in]{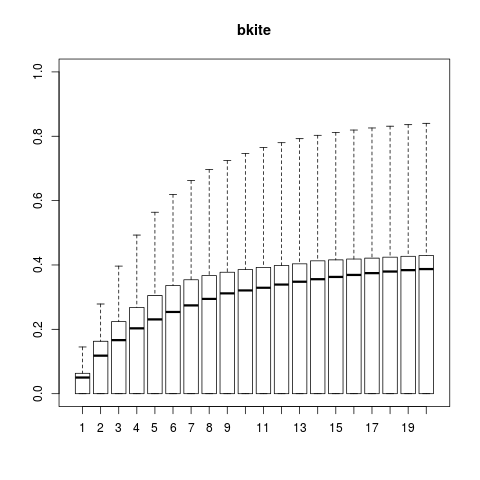}}}
  \subfigure[V  = 60\%, Q=35\%]{ \label{fig:deg:c4}
       {\includegraphics[width=1.24in]{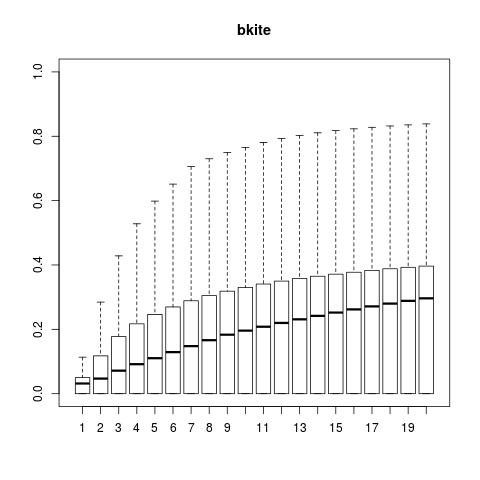}}}
  \subfigure[V  = 60\%, Q=45\%]{ \label{fig:deg:c5}
    {\includegraphics[width=1.24in]{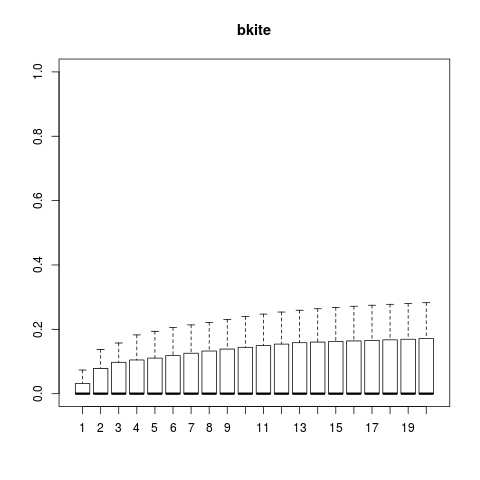}}}\\
  \subfigure[V  = 85\%, Q=05\%]{ \label{fig:deg:d1}
       {\includegraphics[width=1.24in]{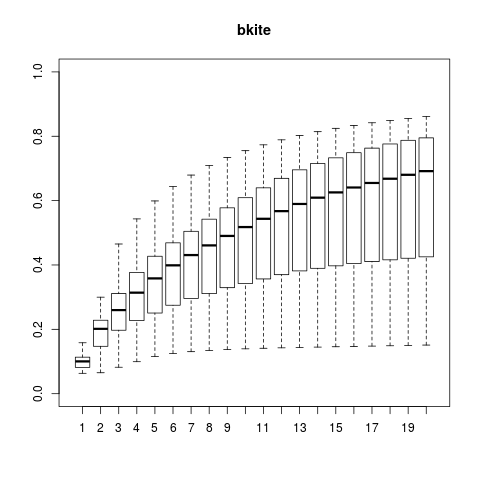}}}
  \subfigure[V  = 85\%, Q=15\%]{ \label{fig:deg:d2}
       {\includegraphics[width=1.24in]{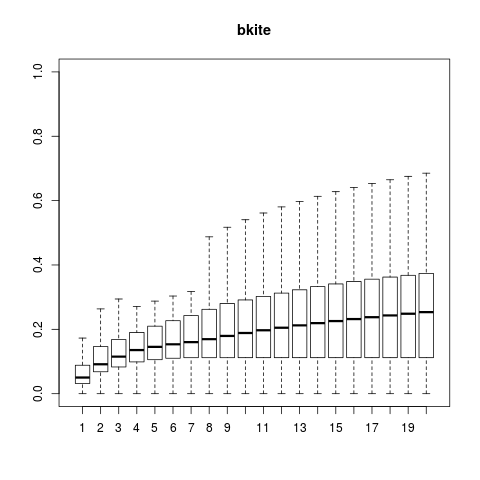}}}
  \subfigure[V  = 85\%, Q=25\%]{ \label{fig:deg:d3}
       {\includegraphics[width=1.24in]{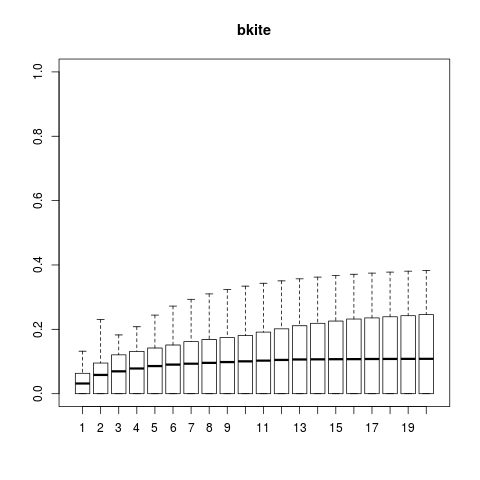}}}
  \subfigure[V  = 85\%, Q=35\%]{ \label{fig:deg:d4}
       {\includegraphics[width=1.24in]{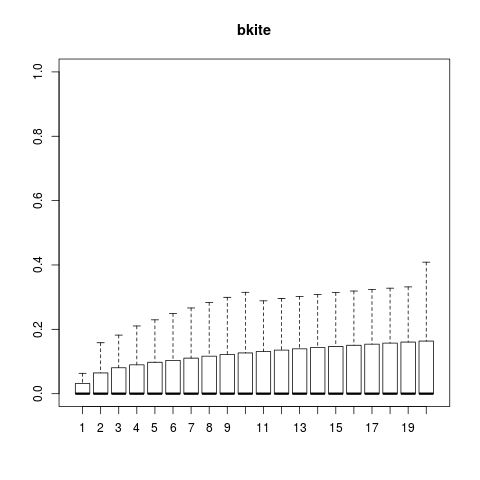}}}
  \subfigure[V  = 85\%, Q=45\%]{ \label{fig:deg:d5}
    {\includegraphics[width=1.24in]{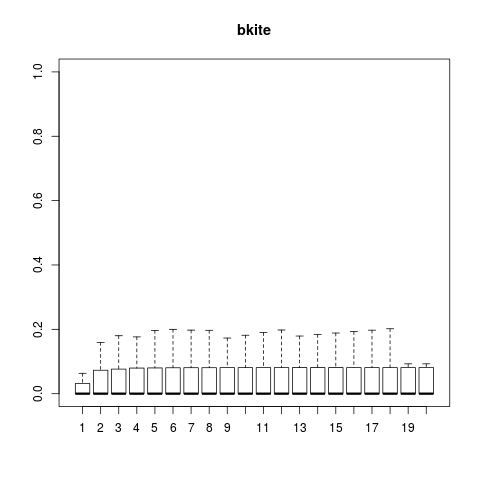}}}
  \caption{Effectiveness of super-spreader suppression in Brightkite
    graph. $V$ is the fraction of the contact graph visible to the
    responder. $Q$ is the fraction of the population asked to
    self-isolate.  Even at 10\% visibility, propagation is
    substantially reduced. Further increase in visibility does not
    bring additional benefits. Notably, propagation minimisation is
    not achieved even with high isolation budgets (unlike super-link
    isolation). Super-Spreaders are easy to spot but suffers from the
    need for high isolation budgets. Full suppression of disease
    transmission not achieved (unlike super-link strategy) even at the
    highest visibility and isolation-budget allocations.}
  \label{fig:deg}
\end{figure*}

Figure~\ref{fig:deg} shows the amortised results over 10000 randomised
simulations. We can observe that proactive isolation of
super-spreaders is somewhat effective in limiting disease propagation,
and it does significantly better than contact tracing, but not as good as
super-link detection. Isolation budgets plays an important role
compared to visibility. Overall we can observe that moderate
visibility with high isolation budgets result in flattening the
transmission potential of the network from 0.8 to 0.2, which is better
than contact tracing but not as good as super-link isolation, which
consistently achieves the lower bound of transmission potential
(maximal curve-flattening).


%

\section{Discussion}




\paragraphb{Topology of disease transmission:}
The topology of the proximity graph plays an important role in the
rate of disease propagation. Our results indicate that topology has
some influence on the ability to stop a disease outbreak. Our work may
pave the way for developing efficient strategies for increasing
resilience against an epidemic within a given population. As it turns
out, contact tracing is useful but it is possible to do much better --
while contract tracing only works reasonably well once the visibility
is greater than 60\%, super-link detection is as effective as contact
tracing at just 10\% visibility. This is a significant improvement
making graph-based approaches practical.

\paragraphb{Collaboration at many scales:}
Our experiments show that it is possible for multiple entities to
safely collaborate in defending populations against an epidemic, and
in the presence of partial observations. These entities can be as
small as individuals (micro-collaboration); or, a regional or national
health agency (macro-collaboration); and, still work with each other
(mixed-collaboration). However, several practical issues remain. For
best results, macro-collaborations are likely to lead to better
outcomes over micro- or mixed-collaborations due to the economics of
distributed disease propagation analysis. This is because a regional
or national health agency working together may result in a better
overall perspective whereas the threshold of the number of individuals
working together will be relatively higher and the one-off costs
involved will result in inefficiencies. This calls for efficient
protocols for privacy-preserving super-link detection in
micro-collaborations. Aside from economics, the security assumption of
semi-honest adversaries lends support to macro-scale collaborations
where such assumptions can be codified into a legally enforceable
contract. This is necessary since the encrypted random-walk data may
reveal sensitive information or violate privacy of individuals.

\paragraphb{Application and containment:} The ability to quickly
identify infections via a comprehensive testing regime is a
pre-requisite for a successful intervention via graph-based approaches
including super-link isolation. Our technique can be seen as a
sophisticated version of its fat-fingered cousin, the full social
lockdown~\cite{das2020critical} -- instead of disabling all proximity
encounters, it is enough to disable a carefully selected fraction as
part of a targeted intervention. It is likely that such a step would
be more socially sustainable as opposed to a full social lockdown. It
is also important to note that to be effective, super-link isolation
must be applied pre-emptively before the onset of a wave of infections
(outbreak).

Our intervention approach is symptom agnostic -- it does not consider
whether the person has symptoms or has been in the proximity of an
infected individual. We can see that activities, such as opinion
diffusion and more generally all diffusion processes will also be
slowed or stopped by our techniques. We believe that super-link
detection needs to be combined with appropriate testing to identify
initial sources of infection. A wholesale application of the technique
in the absence of an appropriate testing regime would be wholly
undesirable and counterproductive. Super-Link isolation is not an
societal inoculation strategy nor is it a universal safeguard or
defense against the spread of epidemics. It is at best a carefully
chosen strategy as part of a broader public-health response. To read
more into it would be unsupported by scientific evidence.

\paragraphb{Limits of conventional epidemiological approaches:} A
fundamental challenge within current epidemiological
models~\cite{eichner2003Oxford,ferretti2020Science} of disease
transmission is the assumption that $R_0$, the mean infections per
individual is estimated by a normal distribution. However, real-world
contact graphs have a skewed distribution such as a power-law
distribution~\cite{cho2011kdd}, since social links influence proximity
as well as random chance. Consequently, using $R_0$ to drive a disease
mitigation strategy can result in significant errors. Removing a high
vertex-order node might appear as a good strategy to reduce
$R_0$, however a low vertex-order node may play a bigger role
in disease spreading due to its ability to connect multiple clusters.

In general, the vertex-order or node degree at best has a weak
relationship to disease spreading as vertex-order is a poor indicator
of node importance. Consequently, since $R_0$ is a function of vertex
order, we argue that it is not helpful in designing an efficient
mitigation strategy, although it may be helpful in identifying
thresholds of disease outbreaks.

For measuring the effectiveness of interventions, the transmission
potential (proposed in this paper) is a better metric, as it
recognises that individuals may be uniquely positioned. It is a
function of infection likelihood of all individuals within a
population rather than based on a crude approximation of average
infection rates. It is worth noting that in a power-law distribution
the average is not representative of the typical. As such the
transmission potential is well positioned to capture the influence of
key elements of the propagation network.

\paragraph{The limits of automated contact tracing:}
Contact
tracing~\cite{culnane2020uk,bay2020bluetrace,gov2020privacy,wiertz2020predicted,cdc2020principles,yoneki2012flu,dp3t}
is a commonly applied (or suggested) public-health response in the
face of an infectious disease outbreak. To analyse the effectiveness
of this strategy, we ask the following question provide an analytical
argument: Given a set of partial proximity traces (partly visible
proximity graph), what is the probability that a randomly chosen
disease transmission chain is fully visible to the responder?

A network topology with poor expansion properties (or lower
\emph{eigen-value gap} $\epsilon=1-\lambda_{2}$) tends to have
relatively 'localized' outbreaks, so that, given the first individual
of an infection chain, there exists a subset of individuals within the
network that have a higher chance of being a part of the chain of
transmission than others.

The spectral theory of graphs lends us a few tools, namely chernoff
bounds, in quantifying this risk. Suppose $A$ is the set of
individuals who have installed a contact tracing app, and $\pi_{S}$
the corresponding probability mass of the stationary distribution
$\pi$. The upper bound of the probability that a transmission chain
(random walk) of length $t$ goes through $t_{A}$ nodes of $A$ is given
by Gilbert~\cite{G98}: \( Pr[t_{I}=t] \leq \left(
1+\frac{(1-\pi(I))\epsilon}{10} \right) e^{-t \frac{(1-\pi(I))^2
    \epsilon}{20}}\). However, given that this probability
exponentially decreases with increase in $t$, a small increase in the
length of the transmission chain will negate the benefits of the
contact tracing intervention. As $t$ increases the probability that a
transmission chain is fully visible to responders decreases
exponentially with the length of the chain. The length of the chain is
primarily a function of the incubation period which is approx. 7
days~\cite{backer2020ECDPC}, offering a significant window of
opportunity for the disease to spread. Therefore we argue that contact
tracing is a fairly leaky process which while helpful in slowing down
the spread of the disease can by no means prevent its spread, as empirically demonstrated by our experiments.

\paragraph{Future work:}
As we have demonstrated, disabling super-links can be effective in
slowing down or stopping the transmission of infectious diseases. The
analysis of proximity graphs to identify such elements is a useful
approach towards fighting epidemics. However, many challenges remain
to turn this approach into a full-scale intervention mechanism. We
need to:

\begin{enumerate}
\item Evaluate the scalability of super-link detection on larger
  graphs corresponding to populations of tens of millions.
  
\item Develop efficient set-intersection protocols for the fully
  malicious setting that can support the use case of epidemic
  response.
  
\item Develop mechanisms that incentivise organisations at all scales
  whether public or private to work together in a coordinated response
  to epidemic spreading. Only when incentives combine with efficient
  security protocols can we expect wholesome collaboration at
  multiple scales.

\item Test the dynamic case that involves repeated cycles of disease spread 
  and response.

\item A comparative analysis of the information-theoretic epidemic
  framework we have proposed here with conventional epidemiological
  frameworks.

\end{enumerate}

\section{Related Work}

The impact of the SARS-COV-2 virus is near unprecedented:
organisations closed, and for the first time in 132 years Scottish
examinations were cancelled.  This highlights the need for a more
efficient method for controlling the spread of a virus without
enforcing a 100\% lockdown. Research in the field recognises this,
with a large focus being on contact tracing
mechanisms~\cite{culnane2020uk,bay2020bluetrace,gov2020privacy,wiertz2020predicted,cdc2020principles,yoneki2012flu,dp3t}. Likewise,
some have pursued graph based modeling for justifying contact
tracing~\cite{ferretti2020Science,eichner2003Oxford}. This work builds
on these contributions, to develop novel non-pharmaceutical techniques
based on isolating super-links and mitigating the need for a
nationwide lockdown.

The use of centrality measures to stop disease propagation has been
investigated before. Lee et al.~\cite{lee2013identifying} were the
first to point out the existence of super-spreaders. They showed that
graph centrality metrics can be used to define super-spreading
locations that result in the greatest numbers of infections. They
identify super-spreading cities based on the metrics of degree,
closeness, and betweenness. Instead of nodes at the level of cities,
our focus is on edges at the level of individuals. Specifically, we
find that the isolation of high-conductance elements within a disease
propagation network efficiently prevents an epidemic outbreak, whereas
the isolation of super-spreaders is sub-optimal.

Borgatti~\cite{borgatti2006identifying} explains why: an individual
with low vertex order and low vertex centrality (betweenness
centrality) may still play a key role in disease spreading. Our work
explains the importance of leveraging high-conductance cuts or
super-links in order to prevent an epidemic outbreak. Accurately,
identifying the key players (the super-links) will allow for a focused
use of isolation resources and thus allowing the majority of society
to continue as normal.
 

Colizza et al. work in a similar domain to the super-spreading nature
of cities \cite{colizza2006role}. In their paper, they look at the
air transportation network as a method of identifying and predicting
the spread of infectious disease. They make use of a vertex-edge graph
provided by the Air Transport Association to define a relation between
the pattern of new emergent diseases, and the spread in relation to
the air transport network. This is important for us, as it proves that
graph theory can be used to model, and identify the spread of disease
- this is positive, and shows that the use of vertex-edge graphs may
also be useful to identify super-spreading nodes in a network of
individuals.

Chen et al. highlight in their paper a new and more effective way to
use graphs to reduce the amount of immunization required in a
community \cite{chen2008finding}.  In this paper, as before they
highlight that most current methods make use of degree and betweenness
of graphs to highlight the key player nodes to be removed. They
suggest an alternative method to this by making use of EGP
graphs. This removes the key player nodes, which will be a smaller
number, and this has the impact of a reduction of immunization
dosages, as the key player nodes, creating clusters of localised
groups.  These localised groups are not longer connected due to the
removal of the key player node, which means lesser dosages of
immunizations are required as the virus may not spread to the same
extent of not removing key nodes.

In their paper, Kitsak et al. \cite{kitsak2010identification} look at
using complex networks to identify super spreaders.  They correctly
identify, as other researchers in this domain do, degree isn't
necessarily the deciding factor of how spreading will occur in a
network. The focus is on the ability of where in the network an
infected node can reach to.  That is, if there are individual clusters
on the node, which are separated, and only one cluster is infected,
then it's not possible for the other uninfected clusters to become
infected as there's no edge (interaction) for the infection to travel
across. In this paper, they offer an alternative method of identifying
the spreaders in a graph by using an algorithm which is robust and
able to identify the key spreaders who would cause connectedness in
clusters. This means that those identified spreaders could be
isolated, or avoided, and this would result in an inability of virus
spread throughout the network.

\section{Conclusion}

The past decade has seen tremendous interest and innovation in the
area of disease management. Techniques have been introduced that enable
a variety of non-clinical interventions to halt or slow down the spread
of infectious diseases especially in the instance of a pandemic.
For some time, the scientific community has discussed whether certain
elements within the proximity network bear a disproportionate
responsibility for the spread of an epidemic? While the obvious
candidate has been the notion of super-spreaders --- individuals that
infect many others, the pursuit of super-spreaders has been met with
scepticism on account of their likely role as key workers such as
nurses, bus drivers, and delivery agents, and police personnel. Our
work makes a start on dealing with this question systematically.

We have shown that contact tracing is sub-optimal both analytically
and empirically. Simply tracing infected people, then tracing their
proximity contacts in turn and isolating them is not enough. It slows
disease spreading but leaves the transmission potential high, and
fails to stop the disease under conditions of partial visibility
unless both the visibility (85\%) and the isolation budget (45\%) are
at their peak.

In contrast, equivalent results are obtained via super-link isolation
under low visibility conditions of 10\% at the same isolation
budget. Or, under medium graph visibility of 35\% and a lower
isolation budget of 35\%. Overall, super-link detection is a promising
direction but only one of the many promising avenues of graph-based
approaches to epidemic response.

We have also highlighted the limitations of the $R_0$ metric in
measuring the impact of interventions. The crudeness of $R_0$ is due
to its reliance on vertex-order properties which poorly estimates the
importance of nodes within the propagation graph. We have proposed the
transmission potential metric. Inspired by communications literature,
it measures the capacity of the network to 'carry' infections from one
end to the other and serves as an accurate indicator of effectiveness
of interventions.

\footnotesize
\bibliography{Papers,unstructured-mixes}
\bibliographystyle{abbrv}

\end{document}